\documentclass[aps,prd,preprint,superscriptaddress,amsmath,amssymb,showpacs]{revtex4-1}
\usepackage{dcolumn}
\usepackage{graphicx}
\usepackage{float}
\usepackage{physics}
\usepackage[colorlinks=true,allcolors=blue]{hyperref}

\begin{document}

\title{Heavy quark potential and jet quenching parameter in a rotating D-instanton background}

\author{Jun-Xia Chen}
\email{chenjunxia@mails.ccnu.edu.cn}
\affiliation{Institute of Particle Physics and Key Laboratory of Quark and Lepton Physics (MOS), Central China Normal University,
Wuhan 430079, China}

\author{De-Fu Hou }
\thanks{Corresponding author}
\email{houdf@mail.ccnu.edu.cn}
\affiliation{Institute of Particle Physics and Key Laboratory of Quark and Lepton Physics (MOS), Central China Normal University,
	Wuhan 430079, China}

\date{\today}

\begin{abstract}
	
We get the dual gravity metric of the rotating nuclear matter by performing a standard Lorentz transformation on the static metric in the D-instanton background. Then, we study the effects of the angular velocity, the instanton density and the temperature on the heavy quark potential. It is shown that the angular velocity and the temperature promote dissociation of the quark-antiquark pair, and the instanton density suppresses dissociation. Similarly, according to the result of the jet quenching parameter, we found that the jet quenching parameter increases with the increase of angular velocity, instanton density and temperature, and the jet quenching parameter in the rotating D-instanton background is larger than that of  $\mathcal{N} =4$ SYM theory.
 
\end{abstract}

\maketitle

\section{Introduction}\label{sec:01_intro}

Many efforts have been made to investigate the properties of the quark-gluon plasma (QGP) \cite{SHURYAK200564,ADAMS2005102,ADCOX2005184} since people discovered this new form of matter created in heavy ion collisions. Usually, the free quark and gluon can not be observed because of color confinement. The QGP produced in the laboratory exists for a short time so that only leptons and hadrons are measured by the detectors of the collider. According to the information of the final state particles, we can study the properties of quark-gluon plasma by using QGP probes, such as heavy quarkonium \cite{Maldacena:1998im}, jet \cite{Casalderrey-Solana:2014wca,Wang:1992qdg} and so on.

In the last century, Maldacena conjectured that the strongly coupled gauge theory in four-dimensional spacetime is dual to the weakly coupled string theory on $AdS_{5}\times S^{5}$, which is referred to as AdS/CFT duality \cite{Aharony:1999ti,Maldacena:1998zhr}. According to AdS/CFT duality, many observable quantities can be calculated. In this paper, we will study the heavy quark potential and jet quenching parameter using AdS/CFT. 

The heavy quark potential $V(L)$ increases with the increase of the distance $L$ between the quark-antiquark pair. However, in the context of the QGP, when the distance between the quark-antiquark pair reaches the maximum, the heavy quark potential will no longer increase and is a constant value, which implies that quark-antiquark interaction is screened by the QGP between them. According to the function image between the heavy quark potential and the separation distance, we can know the maximum value of the heavy quark potential and the maximum separation distance of the quark-antiquark pair. The heavy quark potential for $\mathcal{N}=4$ SYM at finite temperature was calculated in \cite{Rey:1998bq,Brandhuber:1998bs}. The subleading term of this observable has been investigated in \cite{Zhang:2011zj}. The influences of different parameters on the heavy quark potential have been studied in many papers. For example, the effect of the hyperscaling violation on the potential was considered in \cite{Zhang:2016jns,Kioumarsipour:2019lvq}. The influence of magnetic fields on the potential has been investigated in \cite{Rougemont:2014efa}. The heavy quark potential as a function of shear viscosity at strong coupling has been discussed in \cite{Noronha:2009ia}. In \cite{Jahnke:2015obr}, the author studied the potential for different values of the Gauss-Bonnet coupling $\lambda_{GB}$. Other important results can be seen in \cite{Andreev:2006ct,Chernicoff:2006hi,Albacete:2008dz,Boschi-Filho:2006hfm,Avramis:2006em,Yang:2015aia,NataAtmaja:2011mfk,Wu:2014gla}.

The jet quenching parameter is defined as the mean transverse momentum acquired by the hard parton per unit distance traveled \cite{Casalderrey-Solana:2011dxg}. According to this parameter, we know how rapidly the energetic parton loses its energy. The first calculation of the jet quenching parameter in hot $\mathcal{N}=4$ supersymmetric QCD was done by H.Liu et al \cite{Liu:2006ug}. They found $\hat{q}_{SYM} = 26.69\sqrt{\alpha_{SYM}N_{c}}T^{3}$ in the large $N_{c}$ and large $\lambda$ limit. In \cite{Armesto:2006zv}, the first correction to the parameter has been computed. After that, the effects of different quantities on the jet quenching parameter have been studied in many works. For instance, the authors investigated the influence of the magnetic fields on the parameter in \cite{Rougemont:2020had,Zhu:2019ujc,PhysRevD.94.085016}. A lattice study of the jet quenching parameter can be seen in \cite{Panero:2013pla}. The parameter in strongly coupled anisotropic $\mathcal{N}=4$ plasma has been studied in \cite{Giataganas:2012zy,Chernicoff:2012gu}. Other related works can be seen in \cite{DEramo:2010wup,BitaghsirFadafan:2010zh,Li:2014hja,Zhang:2019cxu,Liu:2006he,Buchel:2006bv,Nakano:2006js,Zhu:2021nbl}.

An instanton is a solution to equations of motion of the classical field theory on Euclidean spacetime. In order to study the instanton effect from holography, we will work in a D-instanton background, since the D-instanton in string theory is dual to the instanton in super Yang-Mills theory \cite{Park:1998uv,Li:2017ywp}. The D-instanton background is first proposed in \cite{Liu:1999fc}. They consider D-instantons homogeneously distributed over D3-brane, which corresponds to the modification of the pure D3-brane background by adding a RR scalar charge and the dilaton charge. The type IIB string theory in near-horizon D3/D-instanton background is dual to $\mathcal{N} = 4$ SYM theory in a constant self-dual gauge field background. The effect of D-instanton density on different quantities has attracted much interest, such as drag force \cite{Zhang:2018rff}, imaginary potential \cite{Zhang:2017aoc}, Schwinger effect \cite{Shahkarami:2015qff}, and so on. In \cite{Zhang:2016fdk}, the authors have investigated the effect of instanton density on the heavy quark potential and jet quenching parameter. 

Recently, it was found that the fluid produced in non-central collisions has a strong  vortical structure \cite{STAR:2017ckg}. Particles become globally polarized along the direction of rotation because of spin-orbit coupling. According to the results of $ \varLambda $ hyperon polarization measurements, The STAR Collaboration found there was vortex in Au+Au collisions at low energy ($\sqrt{s_{NN}}$ = 7.7-39GeV) \cite{STAR:2017ckg} while no vortex  at $\sqrt{s_{NN}}$ = 62.4GeV and 200GeV \cite{STAR:2007ccu}. The QGP is the most vortical system so far observed, so the effect of vortices on the properties of QGP has attracted intense interest. For example, the influence of rotation on the phase diagram has been studied in \cite{Fujimoto:2021xix,Chernodub:2020qah,Chernodub:2016kxh}. The effect of rotation on the thermodynamic quantities and equations of state (EoS) has been investigated in \cite{Zhou:2021sdy}. The authors used a 5d Kerr-AdS black hole to describe the rotating nuclear matter and calculated the jet quenching parameter in \cite{Golubtsova:2021agl}. Other interesting results can be seen in \cite{McInnes:2018ibt,McInnes:2018mwj,STAR:2018gyt,Ebihara:2016fwa,Becattini:2020ngo,McInnes:2016dwk}.

Inspired by \cite{Zhang:2016fdk,STAR:2017ckg,BravoGaete:2017dso}, we will study the heavy quark potential and jet quenching parameter in a rotating D-instanton background in this paper. To get the metric in a rotating D-instanton background, we should solve the equation of motion of the rotating nuclear matter. However, it is very difficult. We will use the method in \cite{BravoGaete:2017dso,Erices:2017izj,Chen:2020ath} to get a non-conformal rotating black hole solution by performing a standard Lorentz transformation on the static black hole solution. Due to the symmetry of the action, it must be a solution of the Einstein equation. Then, we will calculate the heavy quark potential and jet quenching parameter using the resulting metric and study the effects of angular velocity, D-instanton density and temperature on these two quantities.

The organization of the paper is as follows. In Sec.\ref{sec:02}, we get the metric in a rotating D-instanton background by performing a standard Lorentz transformation on the static metric in the D-instanton background. In Sec.\ref{sec:03}, we study the effects of angular velocity, D-instanton density and temperature on the heavy quark potential. In Sec.\ref{sec:04}, similarly, we study the effects of angular velocity, D-instanton density and temperature on the jet quenching parameter. We conclude with a discussion in Sec.\ref{sec:05}.

\section{Background geometry}\label{sec:02}

Let's briefly review the D-instanton background. We consider D-instantons homogeneously distributed over the D3-brane, which corresponds to modifying the pure D3-brane background by adding a RR scalar charge and the dilaton charge. The ten dimensional super-gravity action in Einstein's frame is \cite{Gibbons:1995vg,Kehagias:1999iy}
\begin{equation}
\label{eq21}
S=\frac{1}{\kappa}\int d^{10}x\sqrt{g}(R-\frac{1}{2}(\partial \Phi)^{2}+\frac{1}{2}e^{2\Phi}(\partial \chi)^{2}-\frac{1}{6}F^{2}_{(5)}),
\end{equation}
where $\Phi$ and $\chi$ denote the dilaton and the axion respectively, and $F_{(5)}$ is a five-form field strength coupled to the D3-branes. If we set $\chi=-e^{-\Phi}+\chi_{0}$, the dilaton term cancel the axion term in Eq.(\ref{eq21}). Then the solution in the string frame is \cite{Li:2021vve,Gwak:2012ht}
\begin{equation}
\label{eq22}
ds^{2}=e^{\frac{\Phi}{2}}[-\frac{r^{2}}{R^{2}}f(r)dt^{2}+\frac{r^{2}}{R^{2}}d\vec{x}^{2}+\frac{1}{f(r)}\frac{R^{2}}{r^{2}}dr^{2}
]+R^{3}d\Omega_{5}^{2},
\end{equation}
with
\begin{equation}
e^{\Phi}=1+\frac{q}{r_{t}^{4}}\log\frac{1}{f(r)},\quad  f(r)=1-\frac{r_{t}^{4}}{r^{4}},\quad q=4\frac {N_D}{N_c}\frac{(2\pi)^4\alpha^{\prime 2}}{V_4}R^4,
\end{equation}
where $(t,\vec{x},r)$ are coordinates of $AdS_{5}$, and $d\Omega_{5}$ denotes the volume element of the five-sphere $S^{5}$. $R$ is the radius of curvature, and $r_{t}$ is the radial position of the event horizon. Gauge theory is on the boundary $r\rightarrow \infty$. The integers $N_D$ and $N_c$ are the D-instanton and D3-branes numbers. $\alpha^\prime$ is related to string tension, and $V_4$ is world volume. The parameter $q$ is associated to the D-instanton density, which also represents the vacuum expectation value of gluon condensation in the dual picture \cite{Shahkarami:2019zax}.

For the sake of calculation, we can convert the metric to cylindrical coordinates,
\begin{equation}
\label{eq23}
	ds^{2}=e^{\frac{\Phi}{2}}[-\frac{r^{2}}{R^{2}}f(r)dt^{2}+\frac{r^{2}}{R^{2}}(dl^{2}+l^{2}d\varphi^{2}+dz^{2})+\frac{1}{f(r)}\frac{R^{2}}{r^{2}}dr^{2}],
\end{equation}	
where $(t,l,\varphi,z,r)$ are coordinates of $AdS_{5}$.

The large vortex will be produced if the heavy-ion collision is a non-central collision. In order to investigate the effect of the rotation on the QGP, we should calculate the observables in the rotating background. It is very difficult to solve the equation of motion of the rotating nuclear matter from the action. Here, we will take an approximate approach. According to \cite{Chen:2020ath,BravoGaete:2017dso,Erices:2017izj}, the metric in the rotating background can be obtained by performing a standard Lorentz transformation on the static metric Eq.(\ref{eq23})
\begin{equation}
\label{eq24}
t\longrightarrow\frac{1}{\sqrt{1-(\omega l)^{2}}}(t+\omega l^{2} \varphi),\quad \varphi\longrightarrow\frac{1}{\sqrt{1-(\omega l)^{2}}}(\varphi+\omega t),
\end{equation}
where $\omega$ is the angular velocity, and $l$ is the radius to the rotating axis. Since we only focus on the qualitative results, we will take $l=1\mathrm{GeV}^{-1}$. The resulting metric is
\begin{equation}
\begin{split}
\label{eq25}
ds^{2}=&e^{\frac{\Phi}{2}}\frac{r^{2}}{R^{2}}[\frac{1}{1-\omega^{2}}(\omega^{2}-f(r))dt^{2}
+\frac{1}{1-\omega^{2}}(1-\omega^{2}f(r))d\varphi^{2}
+\frac{2\omega}{1-\omega^{2}}(1-f(r))dtd\varphi\\&+dl^{2}+dz^{2}+\frac{1}{f(r)}\frac{R^{4}}{r^{4}}dr^{2}].
\end{split}
\end{equation}

In order to calculate  the temperature of the black hole, we need to transform Eq.(\ref{eq25}) into Eq.(\ref{eq26}) below
\begin{equation}
\label{eq26}
ds^2=H(r)(d\varphi+F(r)dt)^{2}+N(r)dt^{2}+e^{\frac{\Phi}{2}}\frac{1}{f(r)}\frac{R^{2}}{r^{2}}dr^{2}
+e^{\frac{\Phi}{2}}\frac{r^{2}}{R^{2}}(dl^{2}+dz^{2}),
\end{equation}
where
\begin{equation}
\begin{split}
\label{eq27}
&H(r)=e^{\frac{\Phi}{2}}\frac{r^{2}}{R^{2}}\frac{1}{1-\omega^{2}}(1-\omega^{2}f(r)),  \\&F(r)=\omega(1-f(r))(1-\omega^{2}f(r))^{-1},
\\&N(r)=e^{\frac{\Phi}{2}}\frac{r^{2}}{R^{2}}(\omega^{2}-1)f(r)(1-\omega^{2}f(r))^{-1}.
\end{split}
\end{equation}

The Hawking temperature of the black hole is given by
\begin{equation}
\label{eq28}
T=|\frac{\lim\limits_{r\to r_{t}}-\frac{1}{2}\sqrt{\frac{g^{rr}}{-\hat{g}_{00}}}\hat{g}_{00,1}}{2\pi}|=\frac{r_{t}\sqrt{1-\omega^{2}}}{\pi R^{2}}.
\end{equation}

Here, $\hat{g}_{00}$ is different from the $t-t$ component of the metric $g_{00}$. $g^{rr}$ is also different from $g_{rr}$.  $\hat{g}_{00,1}$ is the derivative of $\hat{g}_{00}$,
\begin{equation}
	\label{eq35}
	\hat{g}_{00}=N(r),\quad g^{rr}=\frac{1}{g_{rr}}.
\end{equation}

\section{Heavy quark potential}\label{sec:03}

In this section, we study the effects of angular velocity, D-instanton density, and temperature on the heavy quark potential. The quark-antiquark pair is represented by a string whose endpoints are on the D-brane, and the D-brane is at the boundary of AdS. The string which is not straight but goes inside the AdS spacetime is at the lowest energy state. For simplicity, such a string can be approximated by a rectangular string, which can be divided into two parts: the vertical part and the horizontal part. Only the horizontal part contributes to the quark potential. The vertical part gives the quark mass.

The heavy quark potential can be extracted from the expectation value of the Wilson loop. In the limit $\mathcal{T}\rightarrow \infty$, we have
   
\begin{equation}
\label{eq32}
\langle W(C)\rangle\sim e^{-\mathcal{T}V(L)},
\end{equation}
where $C$ is a rectangular loop with one side representing the time $\mathcal{T}$ and the other side representing the separation distance $L$ of the quark-antiquark pair in the AdS spacetime. $V (L)$ represents the heavy quark potential.

In addition, in the large $N_{c}$ and large $\lambda$ limit, we have
\begin{equation}
\label{eq33}
\langle W(C)\rangle\sim e^{-S_{c}}.
\end{equation}

Here, $S_{c}=S-S_{0}$. $S$ represents the Nambu-Goto action of the U-shaped string. $S_{0}$ represents the Nambu-Goto action of the straight string. The Nambu-Goto action is
\begin{equation}
	\label{eq34}
	S=-\frac{1}{2\pi \alpha'}\int d\tau d\sigma \sqrt{-det g_{\alpha \beta}},
\end{equation}
where $\frac{1}{2\pi \alpha'}$ is string tension, $g_{\alpha \beta}$ is known as induced metric
\begin{equation}
	\label{eq35}
	g_{\alpha \beta}=G_{\mu \nu}\frac{\partial X^{\mu}}{\partial \sigma^{\alpha}}\frac{\partial X^{\nu}}{\partial \sigma^{\beta}}.
\end{equation}

Combining the formula (\ref{eq32}) and (\ref{eq33}), we can represent the heavy quark potential in terms of the Nambu-Goto action
\begin{equation}
\label{eq36}
V(L)=\frac{S_{c}}{\mathcal{T}}.
\end{equation}

Now, we are going to calculate the action of the string. A particle draws a world-line in spacetime. A string sweeps a world-sheet in spacetime. We can take the world-sheet coordinates as $\sigma^{\alpha}=(\tau,\sigma)$. Then, the string motion is described by $X^{\mu}(\sigma^{\alpha})$. Here, we take the static gauge 
\begin{equation}
\label{eq37}
\tau=t, \quad \sigma=z,\quad r=r(\sigma).
\end{equation}

Using the world-sheet coordinates $\sigma^{\alpha}$, the $AdS_{5}$ spacetime metric is 
\begin{equation}
\begin{split}
\label{eq38}
ds^{2}&=G_{\mu\nu}dX^{\mu}dX^{\nu}
=g_{\alpha \beta}d\sigma^{\alpha}d\sigma^{\beta}
\\&=e^{\frac{\Phi}{2}}\frac{r^{2}}{R^{2}}[\frac{1}{1-\omega^{2}}(\omega^{2}-f(r))d\tau^{2}+(1+\frac{1}{f(r)}\frac{R^{4}}{r^{4}}\dot{r}^{2})d\sigma^{2}],
\end{split}
\end{equation}
then, the action is given by
\begin{equation}
\label{eq39}
S=\frac{\mathcal{T}}{2\pi \alpha'}\int d\sigma \sqrt{e^{\Phi}\frac{1}{1-\omega^{2}}(f(r)-\omega^{2})(\frac{\dot{r}^{2}}{f(r)}+\frac{r^{4}}{R^{4}})},
\end{equation}
where $\dot{r}=\frac{dr}{d\sigma}$, $\mathcal{T}$ is the time duration in $t$. According to the action, the Lagrangian does not contain $\sigma$, so we have a conserved quantity
\begin{equation}
\label{eq311}
\frac{\partial \mathcal{L}}{\partial \dot{r}}\dot{r}-\mathcal{L}=constant.
\end{equation}

Let us determine the constant. Assuming that the U-shaped string has the turning point at $\sigma=0, r=r_{c}$, then, $\dot{r}\mid_{r=r_{c}}=0$, so the left of Eq.(\ref{eq311}) becomes a constant when we set $r=r_{c}$
\begin{equation}
\label{eq313}
\frac{e^{\Phi}\frac{1}{1-\omega^{2}}(f(r)-\omega^{2})\frac{r^{4}}{R^{4}}}
{\sqrt{e^{\Phi}\frac{1}{1-\omega^{2}}(f(r)-\omega^{2})(\frac{\dot{r}^{2}}{f(r)}+\frac{r^{4}}{R^{4}})}}
=\sqrt{e^{\Phi(r_{c})}\frac{1}{1-\omega^{2}}(f(r_{c})-\omega^{2})\frac{r_{c}^{4}}{R^{4}}}.
\end{equation}

\begin{figure}
	\centering
	\includegraphics[scale=0.4]{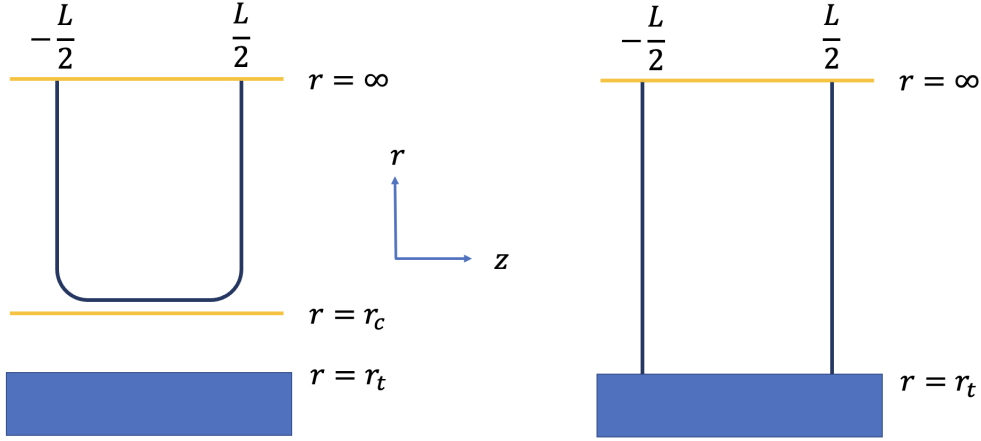}
	\caption{The left is U-shaped string, the right is straight string.}
	\label{string}
\end{figure}	

After a simple calculation, we have
\begin{equation}
\label{eq314}
	\dot{r}=\frac{r^{2}}{R^{2}}
\sqrt{f(r)\frac{g(r)-g(r_{c})}{g(r_{c})}},
\end{equation}
where
\begin{equation}
\begin{split}
		\label{eq3141}
		&g(r)=e^{\Phi}(f(r)-\omega^{2})r^{4}, \quad g(r_{c})=e^{\Phi(r_{c})}(f(r_{c})-\omega^{2})r_{c}^{4},
		\\&e^{\Phi(r_{c})}=1+\frac{q}{r_{t}^{4}}\log\frac{1}{f(r_{c})},\quad  f(r_{c})=1-\frac{r_{t}^{4}}{r_{c}^{4}}.
\end{split}
\end{equation}

As shown in Fig.\ref{string}, the quark and antiquark are located at $z=\frac{L}{2}, z=-\frac{L}{2}$ respectively if the distance between the quark-antiquark pair is $L$. Integrating the above equation, we have
\begin{equation}
\label{eq315}
L=2R^{2}\int_{r_{c}}^{\infty} dr\sqrt{\frac{g(r_{c})}{r^4f(r)[g(r)-g(r_{c})]}},
\end{equation}

\begin{figure}
	\centering
	\includegraphics[scale=0.41]{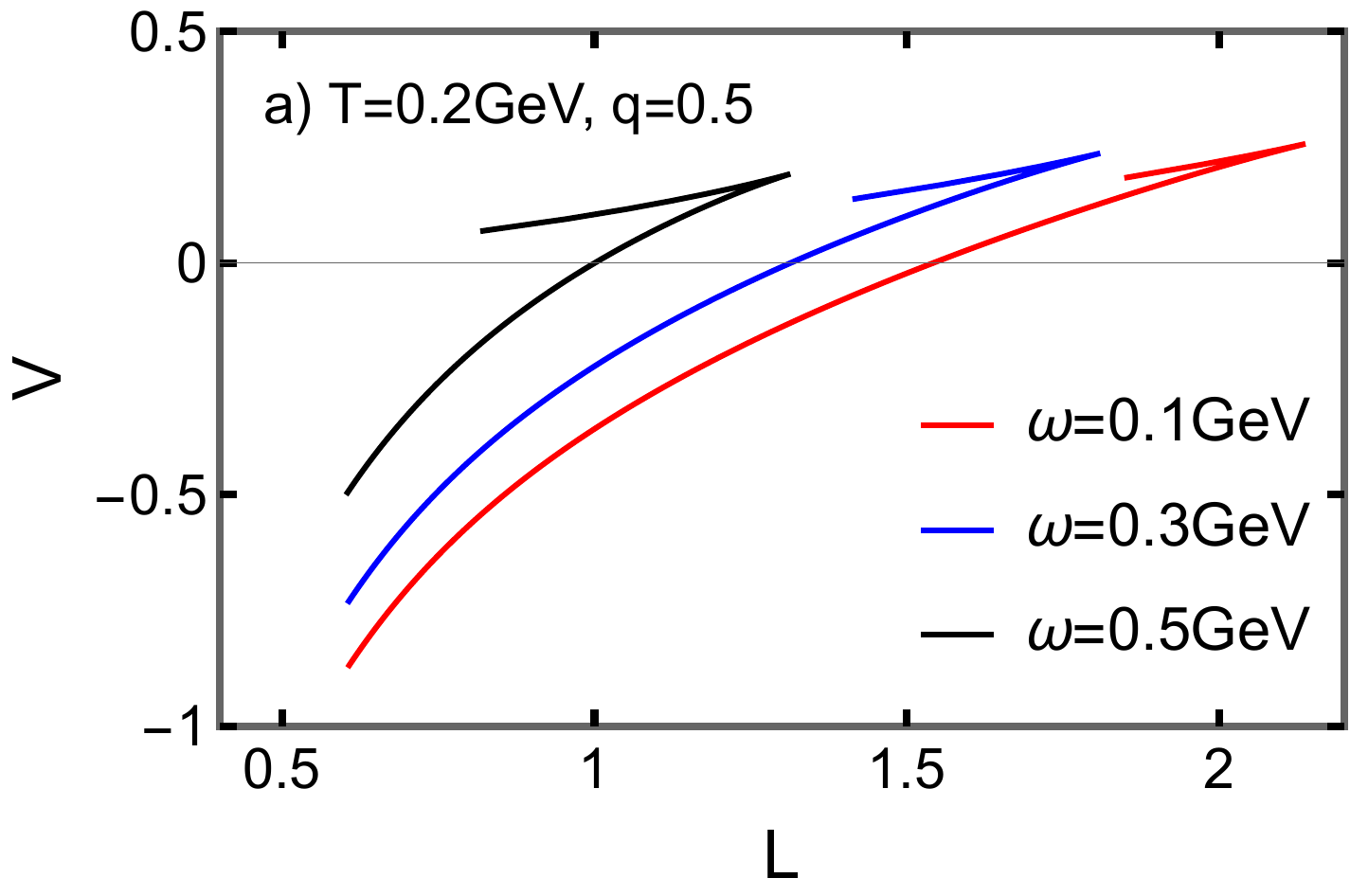}
	\includegraphics[scale=0.41]{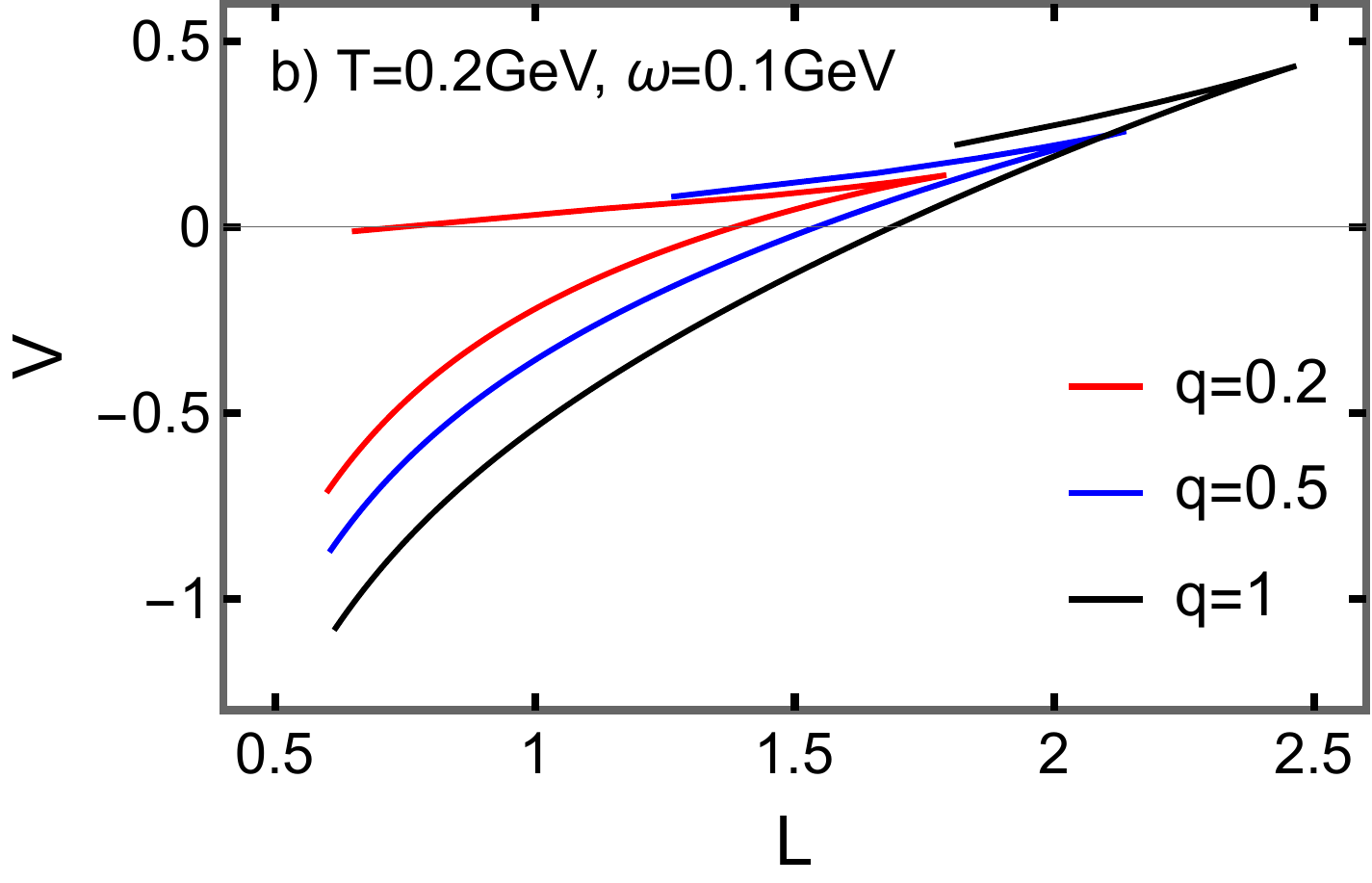}
	\includegraphics[scale=0.41]{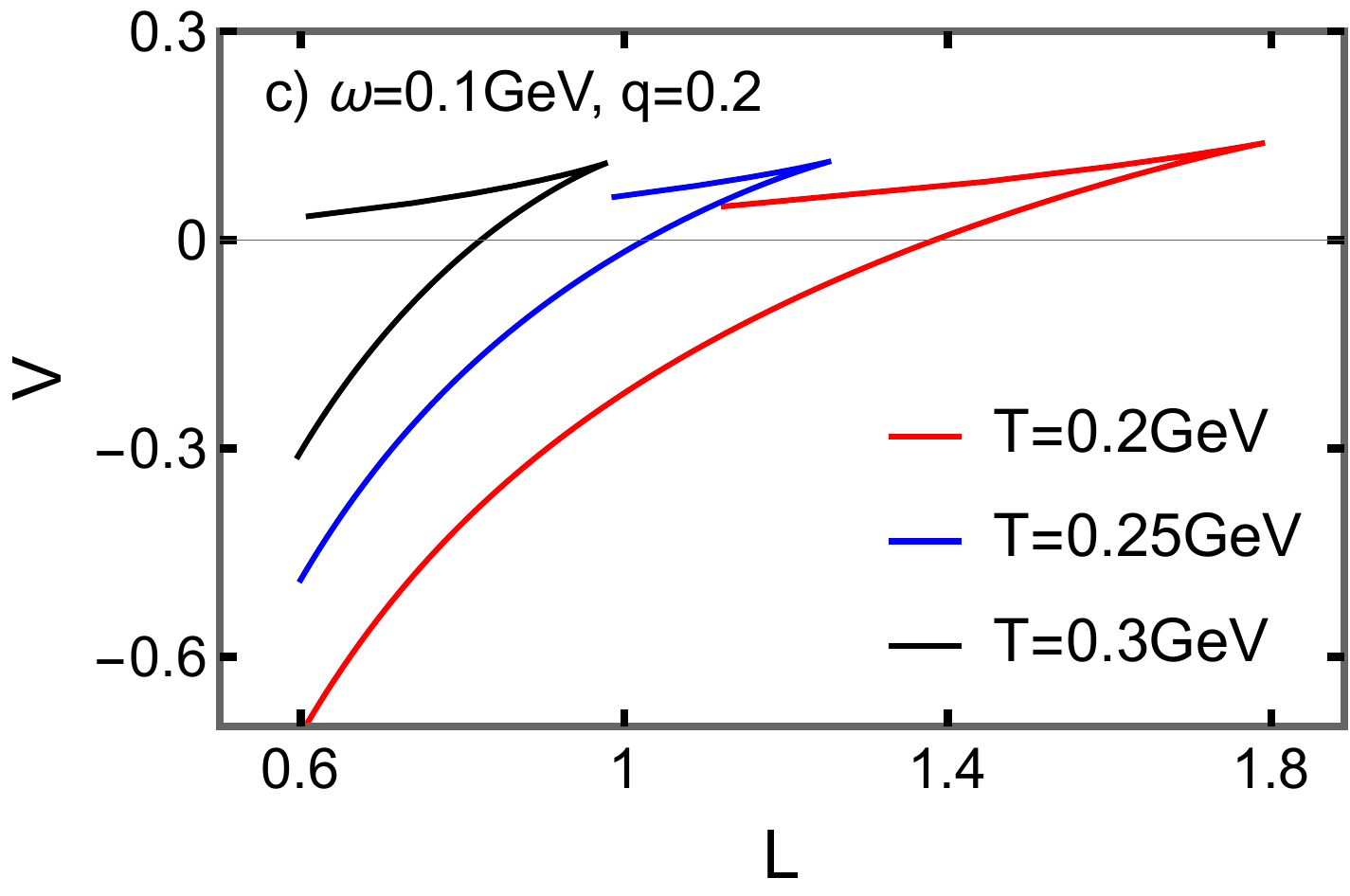}	
	\caption{Heavy quark potential $V(L)$ as a function of distance between the quark-antiquark pair. (a)for different angular velocities, (b)for different parameter $q$, (c)for different temperatures.}
	\label{vl}
\end{figure}

the unit of $L$ is $\mathrm{GeV}^{-1}$. Combining Eq.(\ref{eq39}) and Eq.(\ref{eq314}), the action of U-shaped string is
\begin{equation}
\label{eq316}
S=\frac{\mathcal{T}}{\pi \alpha'}\int_{r_{c}}^{\infty}dr
\sqrt{\frac{e^{\Phi}(f(r)-\omega^{2})g(r)}
	{(1-\omega^{2})f(r)[g(r)-g(r_{c})]}},
\end{equation}
it represents the total energy of the heavy quarkonium. We need to subtract the masses of the quark and antiquark to get the potential between the quark-antiquark pair. The action of the straight string connecting the boundary and the horizon represents the mass of the quark 
\begin{equation}
\label{eq317}
S_{0}=\frac{\mathcal{T}}{\pi \alpha'}\int_{r_{t}}^{\infty}dr\sqrt{-g_{tt}g_{rr}}
=\frac{\mathcal{T}}{\pi \alpha'}\int_{r_{t}}^{\infty}dr\sqrt{\frac{e^{\Phi}(f(r)-\omega^{2})}{f(r)(1-\omega^{2})}}.
\end{equation}

The  heavy quark potential is
\begin{equation}
\label{eq319}
V(L)=\frac{1}{\pi \alpha'}\{\int_{r_{c}}^{\infty}dr
\sqrt{\frac{e^{\Phi}(f(r)-\omega^{2})g(r)}
	{(1-\omega^{2})f(r)[g(r)-g(r_{c})]}}
-\int_{r_{t}}^{\infty}dr\sqrt{\frac{e^{\Phi}(f(r)-\omega^{2})}{f(r)(1-\omega^{2})}}\}.
\end{equation}

The result is the same as \cite{Zhang:2016fdk} if we set $\omega=0$. Since we just qualitatively analyze the influences of angular velocity, instanton density and temperature on the heavy quark potential, we can set $\frac{1}{\pi \alpha'}=1, R=1$. As shown in Fig.\ref{vl}, we plot the curve of heavy quark potential in terms of the distance between the quark-antiquark pair. There are two parts for each curve: the upper part and the lower part. In fact, only the lower part is significant, the upper part is unphysical. The corresponding $L=L_{m}$ at the turning point of the curve is the maximum separation distance which is also called screening length. At this point, the interaction between quark-antiquark has been screened by QGP. When the distance between the quark-antiquark pair increases further, the heavy quark potential will not increase, that is, when $L\geq L_{m}$, the heavy quark potential is a constant. In the dual geometry, we believe that the bottom of the U-shaped string has reached the horizon at this point.

From Fig.\ref{vl} a), we find that the screening length decreases as the angular velocity increases, and the heavy quark potential increases with the increase of angular velocity at the same separation distance. Both of them indicate that the angular velocity $\omega$ promotes the dissociation of the quarkonium.
From Fig.\ref{vl} b), we find that the screening length increases as the instanton density increases, and the heavy quark potential decreases with the increase of instanton density at the same separation distance. Both of them indicate that the instanton density suppresses dissociation.
From Fig.\ref{vl} c), we find that the screening length decreases as the temperature increases, and the heavy quark potential increases with the increase of temperature at the same separation distance. Both of them indicate that high temperature promotes dissociation.

\section{Jet quenching parameter}\label{sec:04}

In this section, we will calculate the jet quenching parameter in a rotating D-instanton background according to AdS/CFT, and study the influences of angular velocity, D-instanton density, and temperature on the jet quenching parameter. It can be extracted from the expectation value of an adjoint Wilson loop
\begin{equation}
\label{eq41}
\langle W^{A}(\mathcal{C})\rangle \approx \exp[-\frac{1}{4\sqrt{2}}\hat{q}L_{-}L^{2}],
\end{equation}
here, we assume the distance between the quark-antiquark pair is $L$, and they propagate along the light cone through the thermal plasma for a distance $L^{-}$. Contour $\mathcal{C}$ is a rectangular loop, whose one side is along the transverse direction with length $L$ and the other side is in the $x^{-}$ direction with length $L^{-}$.

The expectation value of the lightlike Wilson loop in the fundamental representation refers to the regularized action. According to AdS/CFT correspondence, in the large $N_{c}$ and large $\lambda$ limit, the expectation value is given by 
\begin{equation}
\label{eq42}
\langle W^{F}(\mathcal{C})\rangle \approx \exp[-S_{c}].
\end{equation}

Combining Eq.(\ref{eq41}) and Eq.(\ref{eq42}) and using the relation $\langle W^{A}(\mathcal{C})\rangle\sim \langle W^{F}(\mathcal{C})\rangle^{2}$, we get the expression of the jet quenching parameter
\begin{equation}
\label{eq43}
\hat{q}=8\sqrt{2}\frac{S_{c}}{L_{-}L^{2}}.
\end{equation}

Using light-cone coordinates $x^{\mu}=(r,x^{+},x^{-},z,\varphi)$, the metric Eq.(\ref{eq25}) becomes
\begin{equation}
	\begin{split}
		\label{eq431}
		ds^{2}=&e^{\frac{\Phi}{2}}\frac{r^{2}}{R^{2}}\{\frac{1}{2}\frac{1-f(r)}{1-\omega^{2}}[(dx^{+})^{2}+(dx^{-})^{2}]
		+[\frac{\omega^{2}-f(r)}{1-\omega^{2}}-1]dx^{+}dx^{-}+\frac{1-\omega^{2}f(r)}{1-\omega^{2}}d\varphi^{2}
		\\&+\frac{\sqrt{2}\omega (1-f(r))}{1-\omega^{2}}(dx^{+}+dx^{-})d\varphi+dz^{2}\}
		+e^{\frac{\phi}{2}}\frac{1}{f(r)}\frac{R^{2}}{r^{2}}dr^{2}.
	\end{split}
\end{equation}

We still take the world-sheet coordinates as $\sigma^{\alpha}=(\tau,\sigma)$. The Nambu-Goto action is invariant under coordinate transformation on $\sigma^{\alpha}$, so we can set $\tau=x^{-},\sigma=z$. Here, we assume the quarkonium is along $z$ direction. The Wilson loop is on the surface where $\varphi$ and $x^{+}$ are constant. After gauge fixing, the metric Eq.(\ref{eq431}) becomes
\begin{equation}
\label{eq44}
ds^{2}=e^{\frac{\Phi}{2}}\frac{r^{2}}{R^{2}}[\frac {1}{2}\frac{1-f(r)}{1-\omega^2}d\tau^{2}
+(\frac{1}{f(r)}\frac{R^{4}}{r^{4}}\dot{r}^{2}+1)d\sigma^{2}],
\end{equation}
where $\dot{r}=\frac{dr}{d\sigma}$, using Eq.(\ref{eq34}) and Eq.(\ref{eq35}), the action of the string is 
\begin{equation}
\label{eq45}
S=\frac{\sqrt{2}L_{-}}{2\pi \alpha'}\int_{0}^{\frac{L}{2}}d\sigma \sqrt{e^{\Phi}\frac{1}{1-\omega^{2}}(1-f(r))(\frac{\dot{r}^{2}}{f(r)}+\frac{r^{4}}{R^{4}})}.
\end{equation}

Similarly, the Lagrangian does not contain $\sigma$, we have
\begin{equation}
\label{eq46}
\frac{\partial \mathcal{L}}{\partial \dot{r}}\dot{r}-\mathcal{L}=C.
\end{equation}

Solving the above equation, we get
\begin{equation}
\label{eq47}
\dot{r}=\frac{r^{2}}{R^{2}}\sqrt{f(r)[\frac{1}{C^{2}}e^{\Phi}\frac{1}{1-\omega^{2}}(1-f(r))\frac{r^{4}}{R^{4}}-1]},
\end{equation}
according to the expression of $\dot{r}$, when $f(r)=0$, it is easy to get $\dot{r}=0$. In other words, the turning point of the U-shaped string is at $r=r_{t}$. In the low energy limit ($C\rightarrow 0$), the action of the U-shaped string is
\begin{equation}
\label{eq49}
S=\frac{\sqrt{2}L_{-}}{2\pi \alpha'}\int_{r_{t}}^{\infty}dr
\sqrt{e^{\Phi}\frac{1-f(r)}{(1-\omega^{2})f(r)}}
[1+\frac{1}{2\frac{1}{C^{2}}e^{\Phi}\frac{1}{1-\omega^{2}}(1-f(r))\frac{r^{4}}{R^{4}}}],
\end{equation}
 it represents the total energy of the quarkonium. We need to subtract the energy of the free quark which can be calculated by the action of the straight string connecting the boundary and the horizon
\begin{equation}
\label{eq410}
S_{0}=\frac{2L_{-}}{2\pi \alpha'}\int_{r_{t}}^{\infty} dr \sqrt{g_{--}g_{rr}}
=\frac{\sqrt{2}L_{-}}{2\pi \alpha'}\int_{r_{t}}^{\infty}\sqrt{e^{\Phi}\frac{1-f(r)}{(1-\omega^{2})f(r)}}.
\end{equation}.

\begin{figure}
	\centering
	\includegraphics[scale=0.336]{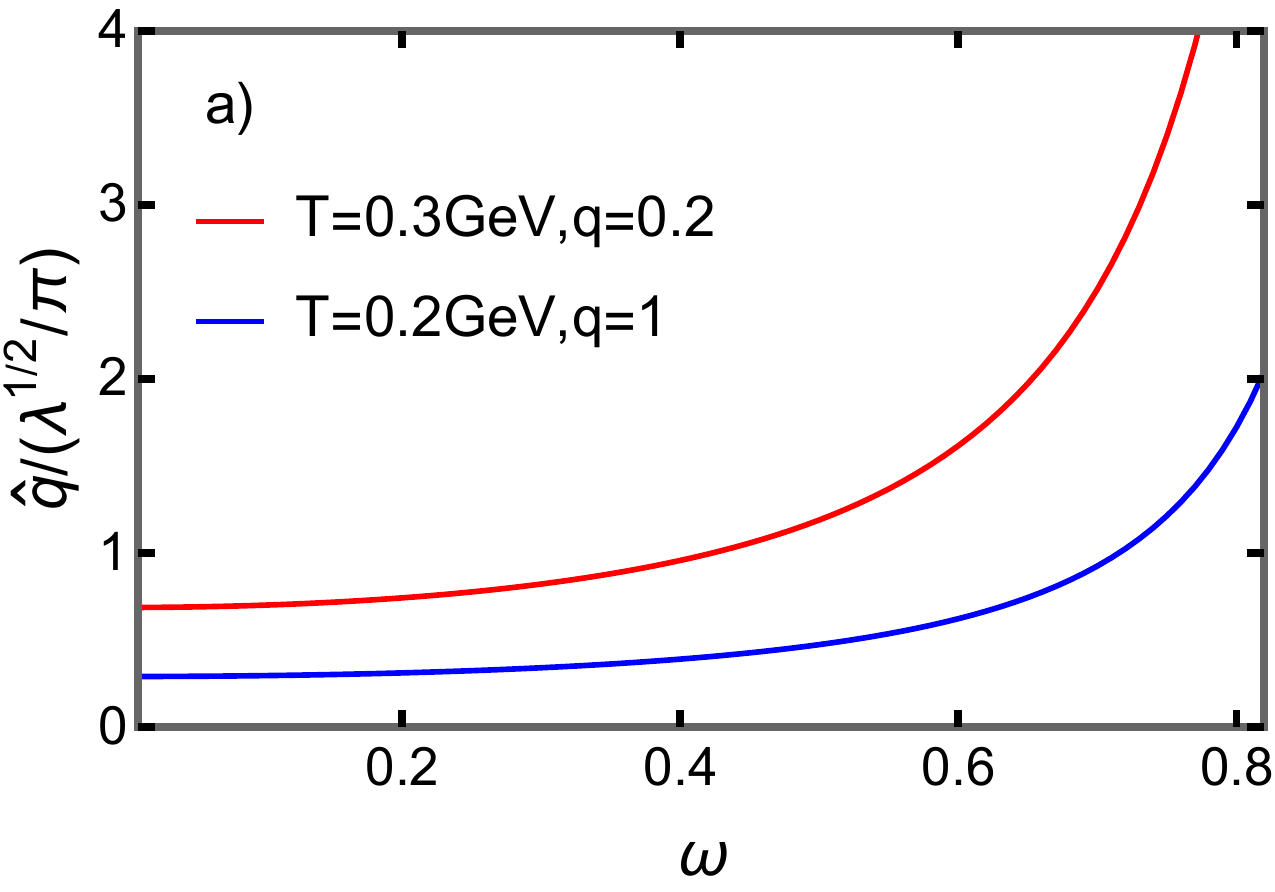}
	\includegraphics[scale=0.3]{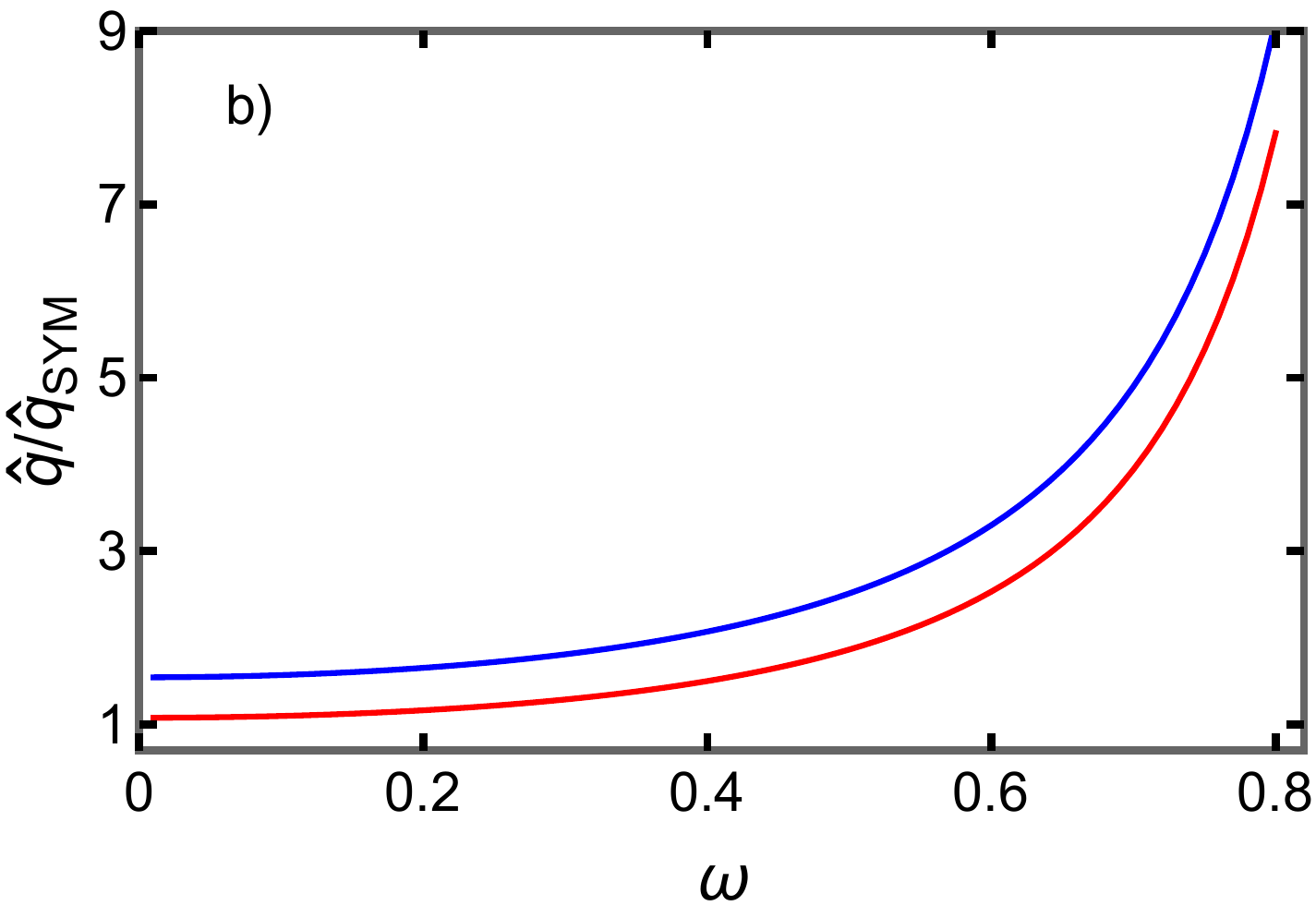}
	\caption{(a) $\hat{q}/\hat{q}_{SYM}$ as a function of angular velocity $\omega$, (b) jet quenching parameter $\hat{q}$ as a function of $\omega$.}
	\label{qw}
\end{figure}

\begin{figure}
	\centering
	\includegraphics[scale=0.336]{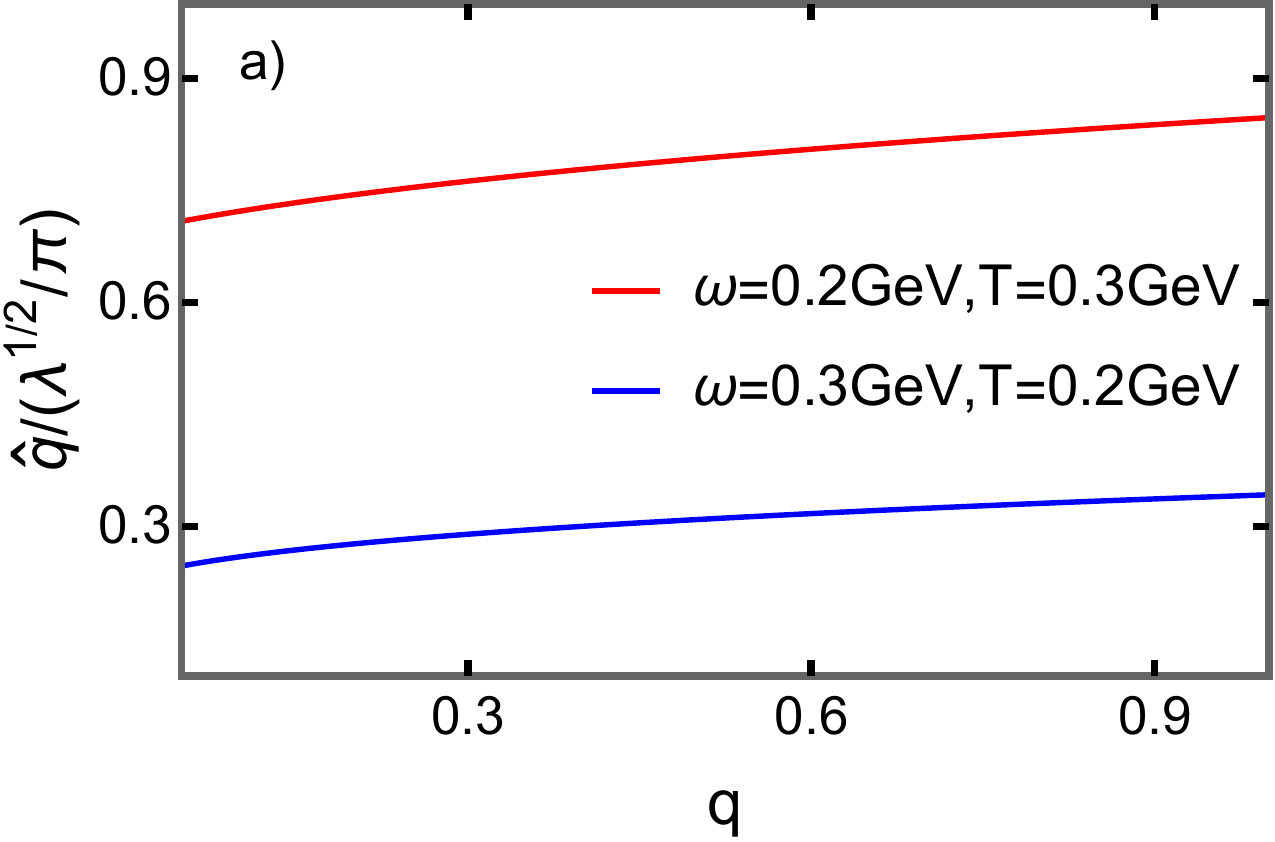}
	\includegraphics[scale=0.3]{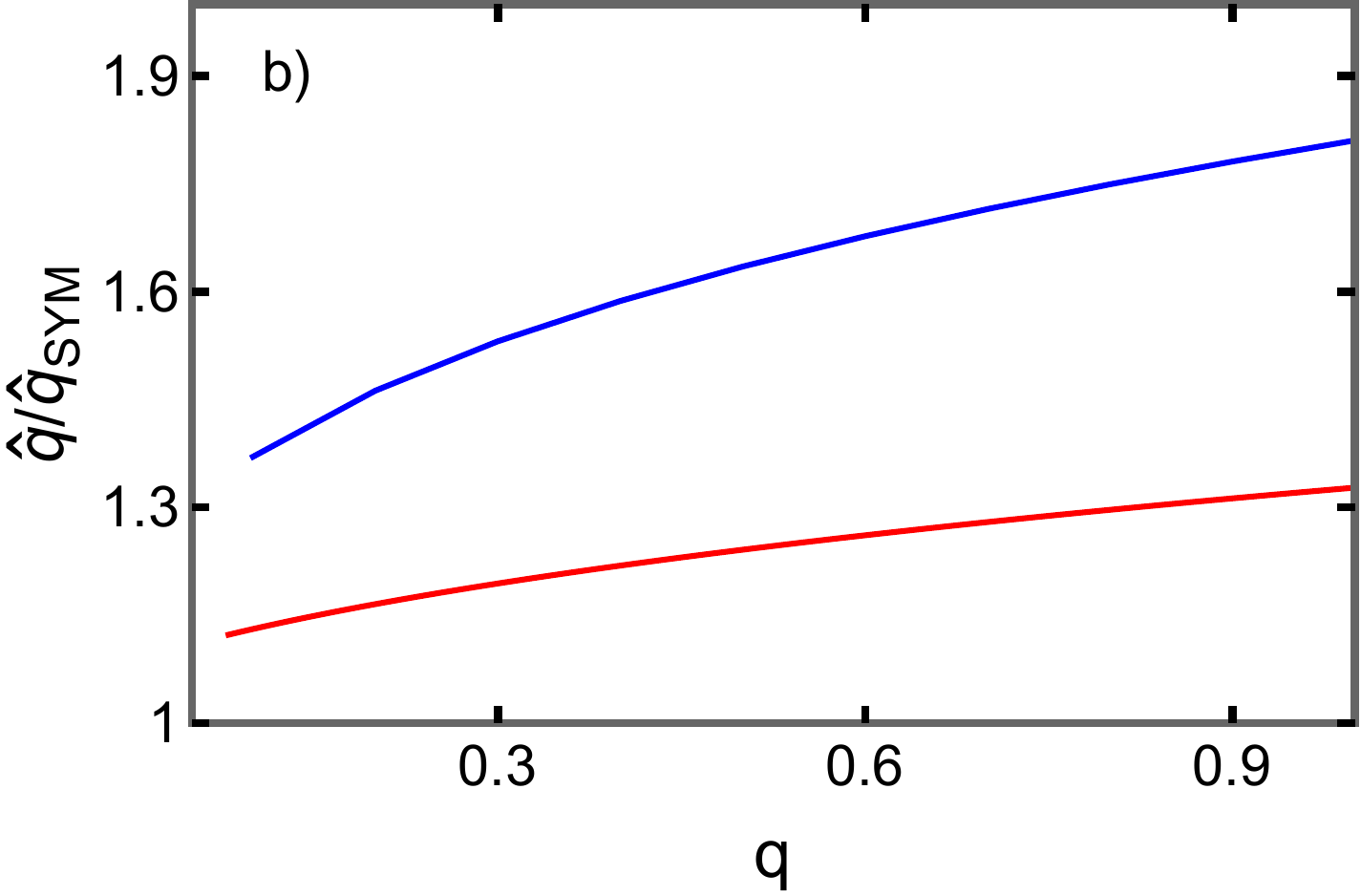}
	\caption{(a) jet quenching parameter $\hat{q}$ as a function of the parameter $q$, (b) $\hat{q}/\hat{q}_{SYM}$ as a function of the parameter $q$.}
	\label{qq}
\end{figure}

\begin{figure}
	\centering
	\includegraphics[scale=0.323]{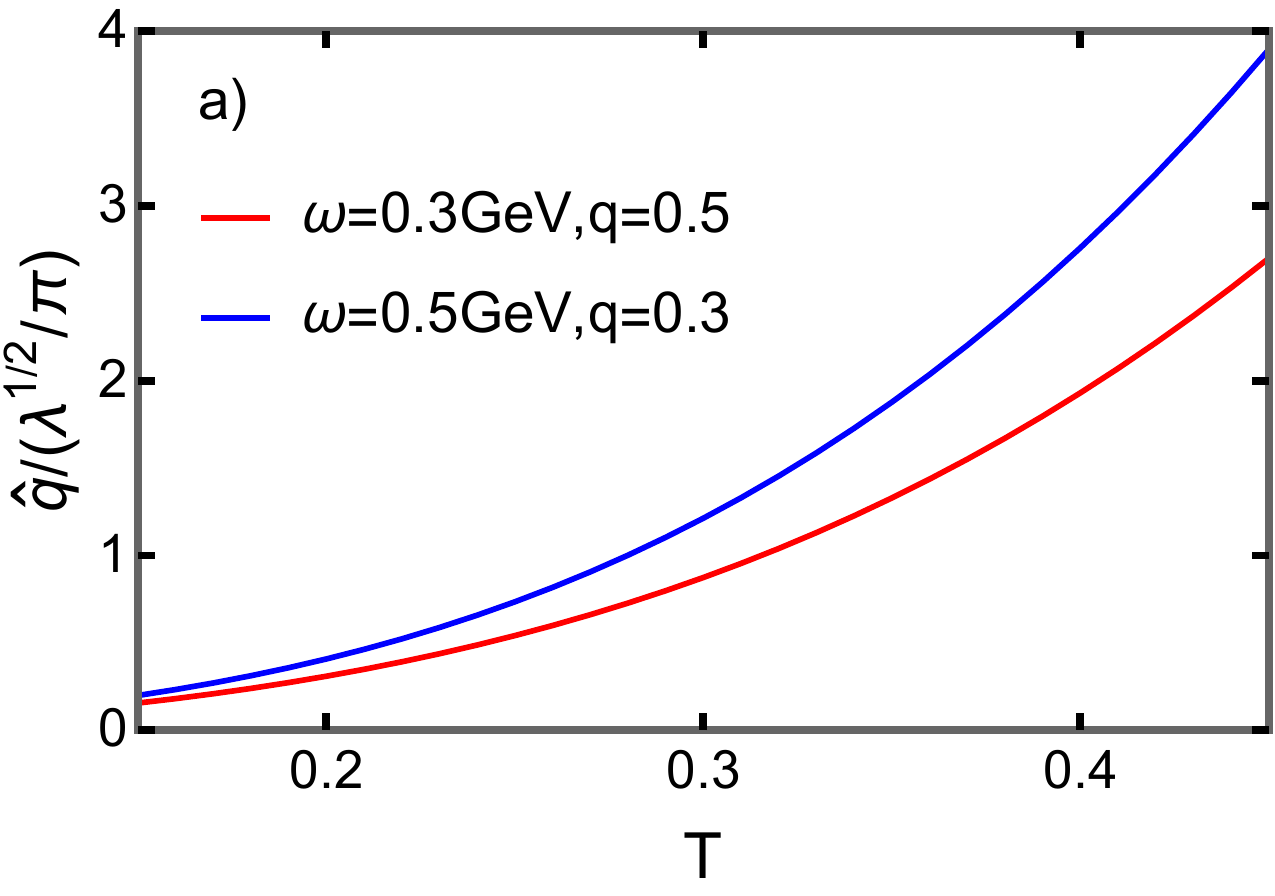}
	\includegraphics[scale=0.3]{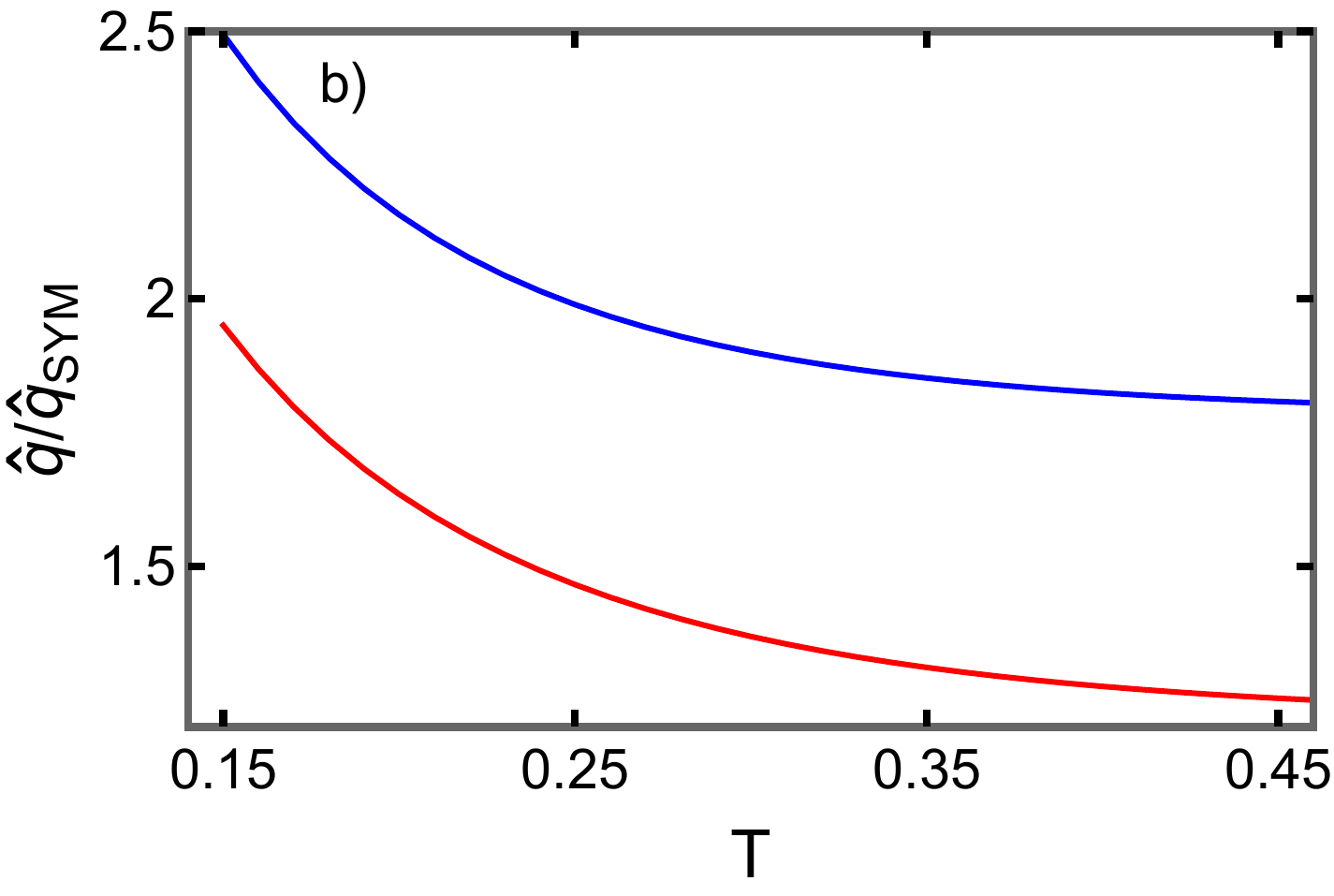}
	\caption{(a) $\hat{q}/\hat{q}_{SYM}$ as a function of the temperature $T$, (b) the jet quenching parameter $\hat{q}$ as a function of the temperature $T$.}
	\label{qt}
\end{figure}

Combining the Eq.(\ref{eq43}), Eq.(\ref{eq49}) and Eq.(\ref{eq410}), the jet quenching parameter is 
\begin{equation}
\label{eq412}
\hat{q}=\frac{1}{\pi \alpha'}\frac{1}{\int_{r_{t}}^{\infty}\sqrt{\frac{1-\omega^{2}}{e^{\Phi}f(r)(1-f(r))}}\frac{R^{4}}{r^{4}}dr}.
\end{equation}

The result is the same as \cite{Zhang:2016fdk} if we set $\omega=0$. We plot the curve of $\hat{q}/\hat{q}_{SYM}$ and $\hat{q}$ in terms of the angular velocity $\omega$, parameter $q$ and temperature $T$ shown separately in Fig.\ref{qw}, Fig.\ref{qq} and Fig.\ref{qt}. Here we set $\frac{1}{\pi \alpha'}=1, R=1$, because we just qualitatively analyze the influences of angular velocity, D-instanton density and temperature on the jet quenching parameter. From Fig.\ref{qw} a), we find the jet quenching parameter increases with the increase of the angular velocity, which indicates that rotation promotes energy loss. From Fig.\ref{qq} a), we find the jet quenching parameter increases with the increase of the D-instanton density, which indicates that the larger the D-instanton density is, the faster the energy loses. From Fig.\ref{qt} a), we find the jet quenching parameter increases with the increase of the temperature, which indicates that high temperature promotes energy loss. From Figure b) of Fig.\ref{qw}, Fig.\ref{qq} and Fig.\ref{qt}, it can be found that the jet quenching parameter in a rotating D-instanton background is larger than that of $\mathcal{N}=4$ SYM theory. It is consistent with the conclusion we obtained according to Figure a) of Fig.\ref{qw}, Fig.\ref{qq} and Fig.\ref{qt}.

To compare the results with the experimental data, we set $\alpha'=0.5$, which is reasonable for temperatures not far above the QCD phase transition \cite{Liu:2006ug}. According to \cite{Gubser:2006qh}, the value of  the ’t Hooft coupling constant is $5.5<\lambda <6\pi$. Here, we set $\lambda=6\pi$, the temperature $T=250$MeV, the parameter $q=1$. From Eq.(\ref{eq412}), we find $\hat{q}=5.4,6.3,9.2\mathrm{GeV^{2}/fm}$ for $\omega=0.1,0.3,0.5$GeV, which is consistent with the RHIC data($5\rightarrow 25\mathrm{GeV^{2}/fm}$) \cite{Edelstein:2008cp}.  

\section{Conclusion and discussion}\label{sec:05}

In this paper, we study the effects of vorticity, D-instanton density and temperature on the properties of QGP. 
Here, two types of QGP probes, heavy quarkonium and jet, are used. To study the influence of the vorticity on the QGP, a coordinate transformation is performed on the static black hole solution to obtain the non-conformal rotating black hole solution. 

According to AdS/CFT, we calculate the heavy quark potential in the rotating D-instanton background and plot the curve of heavy quark potential in terms of the distance between the quark-antiquark pair. We find that the angular velocity and the temperature promote dissociation of the quarkonium, and the D-instanton density suppresses dissociation. Similarly, we also study the effects of angular velocity, D-instanton density, and temperature on the jet quenching parameter, and find that the jet quenching parameter increases with the increase of angular velocity, D-instanton density and temperature, and the jet quenching parameters in the rotating D-instanton background are larger than that of  $\mathcal{N} =4$ SYM theory. Furthermore, the values of the jet quenching parameter we obtained are consistent with the RHIC data when we set $\alpha'=0.5$, $\lambda=6\pi$. 

We expect the effect of vorticity on the dissociation of the quarkonium and heavy quark energy loss we find in this paper can help us to better understand the results of the experiment. The effect of vorticity on the QGP is as important as the influence of the magnetic field, so it is significant to study the effect of vorticity on the other quantities, such as imaginary potential and drag force and so on.    

\section*{Acknowledgments}

This work is supported in part by the National Key Research and Development Program of China under Contract No. 2022YFA1604900. This work is also partly supported by the National Natural Science Foundation of China (NSFC) under Grants No. 12275104, No. 11890711, No. 11890710, and No. 11735007. We would like  to thank Hai-cang Ren, Zi-qiang Zhang and Zhou-run Zhu for useful  discussions.

\section*{References}

\bibliography{ref}

\begin{thebibliography}{71}%
\makeatletter
\providecommand \@ifxundefined [1]{%
 \@ifx{#1\undefined}
}%
\providecommand \@ifnum [1]{%
 \ifnum #1\expandafter \@firstoftwo
 \else \expandafter \@secondoftwo
 \fi
}%
\providecommand \@ifx [1]{%
 \ifx #1\expandafter \@firstoftwo
 \else \expandafter \@secondoftwo
 \fi
}%
\providecommand \natexlab [1]{#1}%
\providecommand \enquote  [1]{``#1''}%
\providecommand \bibnamefont  [1]{#1}%
\providecommand \bibfnamefont [1]{#1}%
\providecommand \citenamefont [1]{#1}%
\providecommand \href@noop [0]{\@secondoftwo}%
\providecommand \href [0]{\begingroup \@sanitize@url \@href}%
\providecommand \@href[1]{\@@startlink{#1}\@@href}%
\providecommand \@@href[1]{\endgroup#1\@@endlink}%
\providecommand \@sanitize@url [0]{\catcode `\\12\catcode `\$12\catcode
  `\&12\catcode `\#12\catcode `\^12\catcode `\_12\catcode `\%12\relax}%
\providecommand \@@startlink[1]{}%
\providecommand \@@endlink[0]{}%
\providecommand \url  [0]{\begingroup\@sanitize@url \@url }%
\providecommand \@url [1]{\endgroup\@href {#1}{\urlprefix }}%
\providecommand \urlprefix  [0]{URL }%
\providecommand \Eprint [0]{\href }%
\providecommand \doibase [0]{http://dx.doi.org/}%
\providecommand \selectlanguage [0]{\@gobble}%
\providecommand \bibinfo  [0]{\@secondoftwo}%
\providecommand \bibfield  [0]{\@secondoftwo}%
\providecommand \translation [1]{[#1]}%
\providecommand \BibitemOpen [0]{}%
\providecommand \bibitemStop [0]{}%
\providecommand \bibitemNoStop [0]{.\EOS\space}%
\providecommand \EOS [0]{\spacefactor3000\relax}%
\providecommand \BibitemShut  [1]{\csname bibitem#1\endcsname}%
\let\auto@bib@innerbib\@empty
\bibitem [{\citenamefont {Shuryak}(2005)}]{SHURYAK200564}%
  \BibitemOpen
  \bibfield  {author} {\bibinfo {author} {\bibfnamefont {E.}~\bibnamefont
  {Shuryak}},\ }\href {\doibase
  https://doi.org/10.1016/j.nuclphysa.2004.10.022} {\bibfield  {journal}
  {\bibinfo  {journal} {Nuclear Physics A}\ }\textbf {\bibinfo {volume}
  {750}},\ \bibinfo {pages} {64} (\bibinfo {year} {2005})}\BibitemShut
  {NoStop}%
\bibitem [{\citenamefont {et~al. (STAR~Collaboration)}(2005)}]{ADAMS2005102}%
  \BibitemOpen
  \bibfield  {author} {\bibinfo {author} {\bibfnamefont {J.~A.}\ \bibnamefont
  {et~al. (STAR~Collaboration)}},\ }\href {\doibase
  https://doi.org/10.1016/j.nuclphysa.2005.03.085} {\bibfield  {journal}
  {\bibinfo  {journal} {Nuclear Physics A}\ }\textbf {\bibinfo {volume}
  {757}},\ \bibinfo {pages} {102} (\bibinfo {year} {2005})}\BibitemShut
  {NoStop}%
\bibitem [{\citenamefont {et~al. (PHENIX~Collaboration)}(2005)}]{ADCOX2005184}%
  \BibitemOpen
  \bibfield  {author} {\bibinfo {author} {\bibfnamefont {K.~A.}\ \bibnamefont
  {et~al. (PHENIX~Collaboration)}},\ }\href {\doibase
  https://doi.org/10.1016/j.nuclphysa.2005.03.086} {\bibfield  {journal}
  {\bibinfo  {journal} {Nuclear Physics A}\ }\textbf {\bibinfo {volume}
  {757}},\ \bibinfo {pages} {184} (\bibinfo {year} {2005})}\BibitemShut
  {NoStop}%
\bibitem [{\citenamefont {Maldacena}(1998)}]{Maldacena:1998im}%
  \BibitemOpen
  \bibfield  {author} {\bibinfo {author} {\bibfnamefont {J.~M.}\ \bibnamefont
  {Maldacena}},\ }\href {\doibase 10.1103/PhysRevLett.80.4859} {\bibfield
  {journal} {\bibinfo  {journal} {Phys. Rev. Lett.}\ }\textbf {\bibinfo
  {volume} {80}},\ \bibinfo {pages} {4859} (\bibinfo {year} {1998})},\ \Eprint
  {http://arxiv.org/abs/hep-th/9803002} {arXiv:hep-th/9803002} \BibitemShut
  {NoStop}%
\bibitem [{\citenamefont {Casalderrey-Solana}\ \emph
  {et~al.}(2014{\natexlab{a}})\citenamefont {Casalderrey-Solana}, \citenamefont
  {Gulhan}, \citenamefont {Milhano}, \citenamefont {Pablos},\ and\
  \citenamefont {Rajagopal}}]{Casalderrey-Solana:2014wca}%
  \BibitemOpen
  \bibfield  {author} {\bibinfo {author} {\bibfnamefont {J.}~\bibnamefont
  {Casalderrey-Solana}}, \bibinfo {author} {\bibfnamefont {D.~C.}\ \bibnamefont
  {Gulhan}}, \bibinfo {author} {\bibfnamefont {J.~G.}\ \bibnamefont {Milhano}},
  \bibinfo {author} {\bibfnamefont {D.}~\bibnamefont {Pablos}}, \ and\ \bibinfo
  {author} {\bibfnamefont {K.}~\bibnamefont {Rajagopal}},\ }\href {\doibase
  10.1016/j.nuclphysa.2014.09.019} {\bibfield  {journal} {\bibinfo  {journal}
  {Nucl. Phys. A}\ }\textbf {\bibinfo {volume} {931}},\ \bibinfo {pages} {487}
  (\bibinfo {year} {2014}{\natexlab{a}})},\ \Eprint
  {http://arxiv.org/abs/1408.5616} {arXiv:1408.5616 [hep-ph]} \BibitemShut
  {NoStop}%
\bibitem [{\citenamefont {Wang}\ and\ \citenamefont
  {Gyulassy}(1992)}]{Wang:1992qdg}%
  \BibitemOpen
  \bibfield  {author} {\bibinfo {author} {\bibfnamefont {X.-N.}\ \bibnamefont
  {Wang}}\ and\ \bibinfo {author} {\bibfnamefont {M.}~\bibnamefont
  {Gyulassy}},\ }\href {\doibase 10.1103/PhysRevLett.68.1480} {\bibfield
  {journal} {\bibinfo  {journal} {Phys. Rev. Lett.}\ }\textbf {\bibinfo
  {volume} {68}},\ \bibinfo {pages} {1480} (\bibinfo {year}
  {1992})}\BibitemShut {NoStop}%
\bibitem [{\citenamefont {Aharony}\ \emph {et~al.}(2000)\citenamefont
  {Aharony}, \citenamefont {Gubser}, \citenamefont {Maldacena}, \citenamefont
  {Ooguri},\ and\ \citenamefont {Oz}}]{Aharony:1999ti}%
  \BibitemOpen
  \bibfield  {author} {\bibinfo {author} {\bibfnamefont {O.}~\bibnamefont
  {Aharony}}, \bibinfo {author} {\bibfnamefont {S.~S.}\ \bibnamefont {Gubser}},
  \bibinfo {author} {\bibfnamefont {J.~M.}\ \bibnamefont {Maldacena}}, \bibinfo
  {author} {\bibfnamefont {H.}~\bibnamefont {Ooguri}}, \ and\ \bibinfo {author}
  {\bibfnamefont {Y.}~\bibnamefont {Oz}},\ }\href {\doibase
  10.1016/S0370-1573(99)00083-6} {\bibfield  {journal} {\bibinfo  {journal}
  {Phys. Rept.}\ }\textbf {\bibinfo {volume} {323}},\ \bibinfo {pages} {183}
  (\bibinfo {year} {2000})},\ \Eprint {http://arxiv.org/abs/hep-th/9905111}
  {arXiv:hep-th/9905111} \BibitemShut {NoStop}%
\bibitem [{\citenamefont {Maldacena}(1999)}]{Maldacena:1998zhr}%
  \BibitemOpen
  \bibfield  {author} {\bibinfo {author} {\bibfnamefont {J.~M.}\ \bibnamefont
  {Maldacena}},\ }\href {\doibase 10.1063/1.59653} {\bibfield  {journal}
  {\bibinfo  {journal} {AIP Conf. Proc.}\ }\textbf {\bibinfo {volume} {484}},\
  \bibinfo {pages} {51} (\bibinfo {year} {1999})}\BibitemShut {NoStop}%
\bibitem [{\citenamefont {Rey}\ \emph {et~al.}(1998)\citenamefont {Rey},
  \citenamefont {Theisen},\ and\ \citenamefont {Yee}}]{Rey:1998bq}%
  \BibitemOpen
  \bibfield  {author} {\bibinfo {author} {\bibfnamefont {S.-J.}\ \bibnamefont
  {Rey}}, \bibinfo {author} {\bibfnamefont {S.}~\bibnamefont {Theisen}}, \ and\
  \bibinfo {author} {\bibfnamefont {J.-T.}\ \bibnamefont {Yee}},\ }\href
  {\doibase 10.1016/S0550-3213(98)00471-4} {\bibfield  {journal} {\bibinfo
  {journal} {Nucl. Phys. B}\ }\textbf {\bibinfo {volume} {527}},\ \bibinfo
  {pages} {171} (\bibinfo {year} {1998})},\ \Eprint
  {http://arxiv.org/abs/hep-th/9803135} {arXiv:hep-th/9803135} \BibitemShut
  {NoStop}%
\bibitem [{\citenamefont {Brandhuber}\ \emph {et~al.}(1998)\citenamefont
  {Brandhuber}, \citenamefont {Itzhaki}, \citenamefont {Sonnenschein},\ and\
  \citenamefont {Yankielowicz}}]{Brandhuber:1998bs}%
  \BibitemOpen
  \bibfield  {author} {\bibinfo {author} {\bibfnamefont {A.}~\bibnamefont
  {Brandhuber}}, \bibinfo {author} {\bibfnamefont {N.}~\bibnamefont {Itzhaki}},
  \bibinfo {author} {\bibfnamefont {J.}~\bibnamefont {Sonnenschein}}, \ and\
  \bibinfo {author} {\bibfnamefont {S.}~\bibnamefont {Yankielowicz}},\ }\href
  {\doibase 10.1016/S0370-2693(98)00730-8} {\bibfield  {journal} {\bibinfo
  {journal} {Phys. Lett. B}\ }\textbf {\bibinfo {volume} {434}},\ \bibinfo
  {pages} {36} (\bibinfo {year} {1998})},\ \Eprint
  {http://arxiv.org/abs/hep-th/9803137} {arXiv:hep-th/9803137} \BibitemShut
  {NoStop}%
\bibitem [{\citenamefont {Zhang}\ \emph {et~al.}(2011)\citenamefont {Zhang},
  \citenamefont {Hou}, \citenamefont {Ren},\ and\ \citenamefont
  {Yin}}]{Zhang:2011zj}%
  \BibitemOpen
  \bibfield  {author} {\bibinfo {author} {\bibfnamefont {Z.-q.}\ \bibnamefont
  {Zhang}}, \bibinfo {author} {\bibfnamefont {D.}~\bibnamefont {Hou}}, \bibinfo
  {author} {\bibfnamefont {H.-c.}\ \bibnamefont {Ren}}, \ and\ \bibinfo
  {author} {\bibfnamefont {L.}~\bibnamefont {Yin}},\ }\href {\doibase
  10.1007/JHEP07(2011)035} {\bibfield  {journal} {\bibinfo  {journal} {JHEP}\
  }\textbf {\bibinfo {volume} {07}},\ \bibinfo {pages} {035} (\bibinfo {year}
  {2011})},\ \Eprint {http://arxiv.org/abs/1104.1344} {arXiv:1104.1344
  [hep-ph]} \BibitemShut {NoStop}%
\bibitem [{\citenamefont {Zhang}\ \emph
  {et~al.}(2017{\natexlab{a}})\citenamefont {Zhang}, \citenamefont {Ma},
  \citenamefont {Hou},\ and\ \citenamefont {Chen}}]{Zhang:2016jns}%
  \BibitemOpen
  \bibfield  {author} {\bibinfo {author} {\bibfnamefont {Z.-q.}\ \bibnamefont
  {Zhang}}, \bibinfo {author} {\bibfnamefont {C.}~\bibnamefont {Ma}}, \bibinfo
  {author} {\bibfnamefont {D.-f.}\ \bibnamefont {Hou}}, \ and\ \bibinfo
  {author} {\bibfnamefont {G.}~\bibnamefont {Chen}},\ }\href {\doibase
  10.1155/2017/8276534} {\bibfield  {journal} {\bibinfo  {journal} {Adv. High
  Energy Phys.}\ }\textbf {\bibinfo {volume} {2017}},\ \bibinfo {pages}
  {8276534} (\bibinfo {year} {2017}{\natexlab{a}})},\ \Eprint
  {http://arxiv.org/abs/1604.04349} {arXiv:1604.04349 [hep-ph]} \BibitemShut
  {NoStop}%
\bibitem [{\citenamefont {Kioumarsipour}\ and\ \citenamefont
  {Sadeghi}(2019)}]{Kioumarsipour:2019lvq}%
  \BibitemOpen
  \bibfield  {author} {\bibinfo {author} {\bibfnamefont {M.}~\bibnamefont
  {Kioumarsipour}}\ and\ \bibinfo {author} {\bibfnamefont {J.}~\bibnamefont
  {Sadeghi}},\ }\href {\doibase 10.1140/epjc/s10052-019-7132-6} {\bibfield
  {journal} {\bibinfo  {journal} {Eur. Phys. J. C}\ }\textbf {\bibinfo {volume}
  {79}},\ \bibinfo {pages} {636} (\bibinfo {year} {2019})}\BibitemShut
  {NoStop}%
\bibitem [{\citenamefont {Rougemont}\ \emph {et~al.}(2015)\citenamefont
  {Rougemont}, \citenamefont {Critelli},\ and\ \citenamefont
  {Noronha}}]{Rougemont:2014efa}%
  \BibitemOpen
  \bibfield  {author} {\bibinfo {author} {\bibfnamefont {R.}~\bibnamefont
  {Rougemont}}, \bibinfo {author} {\bibfnamefont {R.}~\bibnamefont {Critelli}},
  \ and\ \bibinfo {author} {\bibfnamefont {J.}~\bibnamefont {Noronha}},\ }\href
  {\doibase 10.1103/PhysRevD.91.066001} {\bibfield  {journal} {\bibinfo
  {journal} {Phys. Rev. D}\ }\textbf {\bibinfo {volume} {91}},\ \bibinfo
  {pages} {066001} (\bibinfo {year} {2015})},\ \Eprint
  {http://arxiv.org/abs/1409.0556} {arXiv:1409.0556 [hep-th]} \BibitemShut
  {NoStop}%
\bibitem [{\citenamefont {Noronha}\ and\ \citenamefont
  {Dumitru}(2009)}]{Noronha:2009ia}%
  \BibitemOpen
  \bibfield  {author} {\bibinfo {author} {\bibfnamefont {J.}~\bibnamefont
  {Noronha}}\ and\ \bibinfo {author} {\bibfnamefont {A.}~\bibnamefont
  {Dumitru}},\ }\href {\doibase 10.1103/PhysRevD.80.014007} {\bibfield
  {journal} {\bibinfo  {journal} {Phys. Rev. D}\ }\textbf {\bibinfo {volume}
  {80}},\ \bibinfo {pages} {014007} (\bibinfo {year} {2009})},\ \Eprint
  {http://arxiv.org/abs/0903.2804} {arXiv:0903.2804 [hep-ph]} \BibitemShut
  {NoStop}%
\bibitem [{\citenamefont {Jahnke}\ and\ \citenamefont
  {Misobuchi}(2016)}]{Jahnke:2015obr}%
  \BibitemOpen
  \bibfield  {author} {\bibinfo {author} {\bibfnamefont {V.}~\bibnamefont
  {Jahnke}}\ and\ \bibinfo {author} {\bibfnamefont {A.~S.}\ \bibnamefont
  {Misobuchi}},\ }\href {\doibase 10.1140/epjc/s10052-016-4153-2} {\bibfield
  {journal} {\bibinfo  {journal} {Eur. Phys. J. C}\ }\textbf {\bibinfo {volume}
  {76}},\ \bibinfo {pages} {309} (\bibinfo {year} {2016})},\ \Eprint
  {http://arxiv.org/abs/1510.03774} {arXiv:1510.03774 [hep-th]} \BibitemShut
  {NoStop}%
\bibitem [{\citenamefont {Andreev}\ and\ \citenamefont
  {Zakharov}(2006)}]{Andreev:2006ct}%
  \BibitemOpen
  \bibfield  {author} {\bibinfo {author} {\bibfnamefont {O.}~\bibnamefont
  {Andreev}}\ and\ \bibinfo {author} {\bibfnamefont {V.~I.}\ \bibnamefont
  {Zakharov}},\ }\href {\doibase 10.1103/PhysRevD.74.025023} {\bibfield
  {journal} {\bibinfo  {journal} {Phys. Rev. D}\ }\textbf {\bibinfo {volume}
  {74}},\ \bibinfo {pages} {025023} (\bibinfo {year} {2006})},\ \Eprint
  {http://arxiv.org/abs/hep-ph/0604204} {arXiv:hep-ph/0604204} \BibitemShut
  {NoStop}%
\bibitem [{\citenamefont {Chernicoff}\ \emph {et~al.}(2006)\citenamefont
  {Chernicoff}, \citenamefont {Garcia},\ and\ \citenamefont
  {Guijosa}}]{Chernicoff:2006hi}%
  \BibitemOpen
  \bibfield  {author} {\bibinfo {author} {\bibfnamefont {M.}~\bibnamefont
  {Chernicoff}}, \bibinfo {author} {\bibfnamefont {J.~A.}\ \bibnamefont
  {Garcia}}, \ and\ \bibinfo {author} {\bibfnamefont {A.}~\bibnamefont
  {Guijosa}},\ }\href {\doibase 10.1088/1126-6708/2006/09/068} {\bibfield
  {journal} {\bibinfo  {journal} {JHEP}\ }\textbf {\bibinfo {volume} {09}},\
  \bibinfo {pages} {068} (\bibinfo {year} {2006})},\ \Eprint
  {http://arxiv.org/abs/hep-th/0607089} {arXiv:hep-th/0607089} \BibitemShut
  {NoStop}%
\bibitem [{\citenamefont {Albacete}\ \emph {et~al.}(2008)\citenamefont
  {Albacete}, \citenamefont {Kovchegov},\ and\ \citenamefont
  {Taliotis}}]{Albacete:2008dz}%
  \BibitemOpen
  \bibfield  {author} {\bibinfo {author} {\bibfnamefont {J.~L.}\ \bibnamefont
  {Albacete}}, \bibinfo {author} {\bibfnamefont {Y.~V.}\ \bibnamefont
  {Kovchegov}}, \ and\ \bibinfo {author} {\bibfnamefont {A.}~\bibnamefont
  {Taliotis}},\ }\href {\doibase 10.1103/PhysRevD.78.115007} {\bibfield
  {journal} {\bibinfo  {journal} {Phys. Rev. D}\ }\textbf {\bibinfo {volume}
  {78}},\ \bibinfo {pages} {115007} (\bibinfo {year} {2008})},\ \Eprint
  {http://arxiv.org/abs/0807.4747} {arXiv:0807.4747 [hep-th]} \BibitemShut
  {NoStop}%
\bibitem [{\citenamefont {Boschi-Filho}\ \emph {et~al.}(2006)\citenamefont
  {Boschi-Filho}, \citenamefont {Braga},\ and\ \citenamefont
  {Ferreira}}]{Boschi-Filho:2006hfm}%
  \BibitemOpen
  \bibfield  {author} {\bibinfo {author} {\bibfnamefont {H.}~\bibnamefont
  {Boschi-Filho}}, \bibinfo {author} {\bibfnamefont {N.~R.~F.}\ \bibnamefont
  {Braga}}, \ and\ \bibinfo {author} {\bibfnamefont {C.~N.}\ \bibnamefont
  {Ferreira}},\ }\href {\doibase 10.1103/PhysRevD.74.086001} {\bibfield
  {journal} {\bibinfo  {journal} {Phys. Rev. D}\ }\textbf {\bibinfo {volume}
  {74}},\ \bibinfo {pages} {086001} (\bibinfo {year} {2006})},\ \Eprint
  {http://arxiv.org/abs/hep-th/0607038} {arXiv:hep-th/0607038} \BibitemShut
  {NoStop}%
\bibitem [{\citenamefont {Avramis}\ \emph {et~al.}(2007)\citenamefont
  {Avramis}, \citenamefont {Sfetsos},\ and\ \citenamefont
  {Zoakos}}]{Avramis:2006em}%
  \BibitemOpen
  \bibfield  {author} {\bibinfo {author} {\bibfnamefont {S.~D.}\ \bibnamefont
  {Avramis}}, \bibinfo {author} {\bibfnamefont {K.}~\bibnamefont {Sfetsos}}, \
  and\ \bibinfo {author} {\bibfnamefont {D.}~\bibnamefont {Zoakos}},\ }\href
  {\doibase 10.1103/PhysRevD.75.025009} {\bibfield  {journal} {\bibinfo
  {journal} {Phys. Rev. D}\ }\textbf {\bibinfo {volume} {75}},\ \bibinfo
  {pages} {025009} (\bibinfo {year} {2007})},\ \Eprint
  {http://arxiv.org/abs/hep-th/0609079} {arXiv:hep-th/0609079} \BibitemShut
  {NoStop}%
\bibitem [{\citenamefont {Yang}\ and\ \citenamefont
  {Yuan}(2015)}]{Yang:2015aia}%
  \BibitemOpen
  \bibfield  {author} {\bibinfo {author} {\bibfnamefont {Y.}~\bibnamefont
  {Yang}}\ and\ \bibinfo {author} {\bibfnamefont {P.-H.}\ \bibnamefont
  {Yuan}},\ }\href {\doibase 10.1007/JHEP12(2015)161} {\bibfield  {journal}
  {\bibinfo  {journal} {JHEP}\ }\textbf {\bibinfo {volume} {12}},\ \bibinfo
  {pages} {161} (\bibinfo {year} {2015})},\ \Eprint
  {http://arxiv.org/abs/1506.05930} {arXiv:1506.05930 [hep-th]} \BibitemShut
  {NoStop}%
\bibitem [{\citenamefont {Nata~Atmaja}\ \emph {et~al.}(2015)\citenamefont
  {Nata~Atmaja}, \citenamefont {Abu~Kassim},\ and\ \citenamefont
  {Yusof}}]{NataAtmaja:2011mfk}%
  \BibitemOpen
  \bibfield  {author} {\bibinfo {author} {\bibfnamefont {A.}~\bibnamefont
  {Nata~Atmaja}}, \bibinfo {author} {\bibfnamefont {H.}~\bibnamefont
  {Abu~Kassim}}, \ and\ \bibinfo {author} {\bibfnamefont {N.}~\bibnamefont
  {Yusof}},\ }\href {\doibase 10.1140/epjc/s10052-015-3795-9} {\bibfield
  {journal} {\bibinfo  {journal} {Eur. Phys. J. C}\ }\textbf {\bibinfo {volume}
  {75}},\ \bibinfo {pages} {565} (\bibinfo {year} {2015})},\ \Eprint
  {http://arxiv.org/abs/1111.7045} {arXiv:1111.7045 [hep-th]} \BibitemShut
  {NoStop}%
\bibitem [{\citenamefont {Wu}\ \emph {et~al.}(2019)\citenamefont {Wu},
  \citenamefont {Hou},\ and\ \citenamefont {Ren}}]{Wu:2014gla}%
  \BibitemOpen
  \bibfield  {author} {\bibinfo {author} {\bibfnamefont {Y.}~\bibnamefont
  {Wu}}, \bibinfo {author} {\bibfnamefont {D.}~\bibnamefont {Hou}}, \ and\
  \bibinfo {author} {\bibfnamefont {H.-c.}\ \bibnamefont {Ren}},\ }\href
  {\doibase 10.1016/j.nuclphysb.2018.11.020} {\bibfield  {journal} {\bibinfo
  {journal} {Nucl. Phys. B}\ }\textbf {\bibinfo {volume} {938}},\ \bibinfo
  {pages} {351} (\bibinfo {year} {2019})},\ \Eprint
  {http://arxiv.org/abs/1401.3635} {arXiv:1401.3635 [hep-ph]} \BibitemShut
  {NoStop}%
\bibitem [{\citenamefont {Casalderrey-Solana}\ \emph
  {et~al.}(2014{\natexlab{b}})\citenamefont {Casalderrey-Solana}, \citenamefont
  {Liu}, \citenamefont {Mateos}, \citenamefont {Rajagopal},\ and\ \citenamefont
  {Wiedemann}}]{Casalderrey-Solana:2011dxg}%
  \BibitemOpen
  \bibfield  {author} {\bibinfo {author} {\bibfnamefont {J.}~\bibnamefont
  {Casalderrey-Solana}}, \bibinfo {author} {\bibfnamefont {H.}~\bibnamefont
  {Liu}}, \bibinfo {author} {\bibfnamefont {D.}~\bibnamefont {Mateos}},
  \bibinfo {author} {\bibfnamefont {K.}~\bibnamefont {Rajagopal}}, \ and\
  \bibinfo {author} {\bibfnamefont {U.~A.}\ \bibnamefont {Wiedemann}},\ }\href
  {\doibase 10.1017/CBO9781139136747} {\emph {\bibinfo {title} {{Gauge/String
  Duality, Hot QCD and Heavy Ion Collisions}}}}\ (\bibinfo  {publisher}
  {Cambridge University Press},\ \bibinfo {year} {2014})\ \Eprint
  {http://arxiv.org/abs/1101.0618} {arXiv:1101.0618 [hep-th]} \BibitemShut
  {NoStop}%
\bibitem [{\citenamefont {Liu}\ \emph {et~al.}(2006)\citenamefont {Liu},
  \citenamefont {Rajagopal},\ and\ \citenamefont {Wiedemann}}]{Liu:2006ug}%
  \BibitemOpen
  \bibfield  {author} {\bibinfo {author} {\bibfnamefont {H.}~\bibnamefont
  {Liu}}, \bibinfo {author} {\bibfnamefont {K.}~\bibnamefont {Rajagopal}}, \
  and\ \bibinfo {author} {\bibfnamefont {U.~A.}\ \bibnamefont {Wiedemann}},\
  }\href {\doibase 10.1103/PhysRevLett.97.182301} {\bibfield  {journal}
  {\bibinfo  {journal} {Phys. Rev. Lett.}\ }\textbf {\bibinfo {volume} {97}},\
  \bibinfo {pages} {182301} (\bibinfo {year} {2006})},\ \Eprint
  {http://arxiv.org/abs/hep-ph/0605178} {arXiv:hep-ph/0605178} \BibitemShut
  {NoStop}%
\bibitem [{\citenamefont {Armesto}\ \emph {et~al.}(2006)\citenamefont
  {Armesto}, \citenamefont {Edelstein},\ and\ \citenamefont
  {Mas}}]{Armesto:2006zv}%
  \BibitemOpen
  \bibfield  {author} {\bibinfo {author} {\bibfnamefont {N.}~\bibnamefont
  {Armesto}}, \bibinfo {author} {\bibfnamefont {J.~D.}\ \bibnamefont
  {Edelstein}}, \ and\ \bibinfo {author} {\bibfnamefont {J.}~\bibnamefont
  {Mas}},\ }\href {\doibase 10.1088/1126-6708/2006/09/039} {\bibfield
  {journal} {\bibinfo  {journal} {JHEP}\ }\textbf {\bibinfo {volume} {09}},\
  \bibinfo {pages} {039} (\bibinfo {year} {2006})},\ \Eprint
  {http://arxiv.org/abs/hep-ph/0606245} {arXiv:hep-ph/0606245} \BibitemShut
  {NoStop}%
\bibitem [{\citenamefont {Rougemont}(2020)}]{Rougemont:2020had}%
  \BibitemOpen
  \bibfield  {author} {\bibinfo {author} {\bibfnamefont {R.}~\bibnamefont
  {Rougemont}},\ }\href {\doibase 10.1103/PhysRevD.102.034009} {\bibfield
  {journal} {\bibinfo  {journal} {Phys. Rev. D}\ }\textbf {\bibinfo {volume}
  {102}},\ \bibinfo {pages} {034009} (\bibinfo {year} {2020})},\ \Eprint
  {http://arxiv.org/abs/2002.06725} {arXiv:2002.06725 [hep-ph]} \BibitemShut
  {NoStop}%
\bibitem [{\citenamefont {Zhu}\ \emph {et~al.}(2019)\citenamefont {Zhu},
  \citenamefont {Feng}, \citenamefont {Shi},\ and\ \citenamefont
  {Zhong}}]{Zhu:2019ujc}%
  \BibitemOpen
  \bibfield  {author} {\bibinfo {author} {\bibfnamefont {Z.-R.}\ \bibnamefont
  {Zhu}}, \bibinfo {author} {\bibfnamefont {S.-Q.}\ \bibnamefont {Feng}},
  \bibinfo {author} {\bibfnamefont {Y.-F.}\ \bibnamefont {Shi}}, \ and\
  \bibinfo {author} {\bibfnamefont {Y.}~\bibnamefont {Zhong}},\ }\href
  {\doibase 10.1103/PhysRevD.99.126001} {\bibfield  {journal} {\bibinfo
  {journal} {Phys. Rev. D}\ }\textbf {\bibinfo {volume} {99}},\ \bibinfo
  {pages} {126001} (\bibinfo {year} {2019})},\ \Eprint
  {http://arxiv.org/abs/1901.09304} {arXiv:1901.09304 [hep-ph]} \BibitemShut
  {NoStop}%
\bibitem [{\citenamefont {Li}\ \emph {et~al.}(2016)\citenamefont {Li},
  \citenamefont {Mamo},\ and\ \citenamefont {Yee}}]{PhysRevD.94.085016}%
  \BibitemOpen
  \bibfield  {author} {\bibinfo {author} {\bibfnamefont {S.}~\bibnamefont
  {Li}}, \bibinfo {author} {\bibfnamefont {K.~A.}\ \bibnamefont {Mamo}}, \ and\
  \bibinfo {author} {\bibfnamefont {H.-U.}\ \bibnamefont {Yee}},\ }\href
  {\doibase 10.1103/PhysRevD.94.085016} {\bibfield  {journal} {\bibinfo
  {journal} {Phys. Rev. D}\ }\textbf {\bibinfo {volume} {94}},\ \bibinfo
  {pages} {085016} (\bibinfo {year} {2016})}\BibitemShut {NoStop}%
\bibitem [{\citenamefont {Panero}\ \emph {et~al.}(2014)\citenamefont {Panero},
  \citenamefont {Rummukainen},\ and\ \citenamefont
  {Sch\"afer}}]{Panero:2013pla}%
  \BibitemOpen
  \bibfield  {author} {\bibinfo {author} {\bibfnamefont {M.}~\bibnamefont
  {Panero}}, \bibinfo {author} {\bibfnamefont {K.}~\bibnamefont {Rummukainen}},
  \ and\ \bibinfo {author} {\bibfnamefont {A.}~\bibnamefont {Sch\"afer}},\
  }\href {\doibase 10.1103/PhysRevLett.112.162001} {\bibfield  {journal}
  {\bibinfo  {journal} {Phys. Rev. Lett.}\ }\textbf {\bibinfo {volume} {112}},\
  \bibinfo {pages} {162001} (\bibinfo {year} {2014})},\ \Eprint
  {http://arxiv.org/abs/1307.5850} {arXiv:1307.5850 [hep-ph]} \BibitemShut
  {NoStop}%
\bibitem [{\citenamefont {Giataganas}(2012)}]{Giataganas:2012zy}%
  \BibitemOpen
  \bibfield  {author} {\bibinfo {author} {\bibfnamefont {D.}~\bibnamefont
  {Giataganas}},\ }\href {\doibase 10.1007/JHEP07(2012)031} {\bibfield
  {journal} {\bibinfo  {journal} {JHEP}\ }\textbf {\bibinfo {volume} {07}},\
  \bibinfo {pages} {031} (\bibinfo {year} {2012})},\ \Eprint
  {http://arxiv.org/abs/1202.4436} {arXiv:1202.4436 [hep-th]} \BibitemShut
  {NoStop}%
\bibitem [{\citenamefont {Chernicoff}\ \emph {et~al.}(2012)\citenamefont
  {Chernicoff}, \citenamefont {Fernandez}, \citenamefont {Mateos},\ and\
  \citenamefont {Trancanelli}}]{Chernicoff:2012gu}%
  \BibitemOpen
  \bibfield  {author} {\bibinfo {author} {\bibfnamefont {M.}~\bibnamefont
  {Chernicoff}}, \bibinfo {author} {\bibfnamefont {D.}~\bibnamefont
  {Fernandez}}, \bibinfo {author} {\bibfnamefont {D.}~\bibnamefont {Mateos}}, \
  and\ \bibinfo {author} {\bibfnamefont {D.}~\bibnamefont {Trancanelli}},\
  }\href {\doibase 10.1007/JHEP08(2012)041} {\bibfield  {journal} {\bibinfo
  {journal} {JHEP}\ }\textbf {\bibinfo {volume} {08}},\ \bibinfo {pages} {041}
  (\bibinfo {year} {2012})},\ \Eprint {http://arxiv.org/abs/1203.0561}
  {arXiv:1203.0561 [hep-th]} \BibitemShut {NoStop}%
\bibitem [{\citenamefont {D'Eramo}\ \emph {et~al.}(2011)\citenamefont
  {D'Eramo}, \citenamefont {Liu},\ and\ \citenamefont
  {Rajagopal}}]{DEramo:2010wup}%
  \BibitemOpen
  \bibfield  {author} {\bibinfo {author} {\bibfnamefont {F.}~\bibnamefont
  {D'Eramo}}, \bibinfo {author} {\bibfnamefont {H.}~\bibnamefont {Liu}}, \ and\
  \bibinfo {author} {\bibfnamefont {K.}~\bibnamefont {Rajagopal}},\ }\href
  {\doibase 10.1103/PhysRevD.84.065015} {\bibfield  {journal} {\bibinfo
  {journal} {Phys. Rev. D}\ }\textbf {\bibinfo {volume} {84}},\ \bibinfo
  {pages} {065015} (\bibinfo {year} {2011})},\ \Eprint
  {http://arxiv.org/abs/1006.1367} {arXiv:1006.1367 [hep-ph]} \BibitemShut
  {NoStop}%
\bibitem [{\citenamefont {Bitaghsir~Fadafan}\ \emph {et~al.}(2011)\citenamefont
  {Bitaghsir~Fadafan}, \citenamefont {Pourhassan},\ and\ \citenamefont
  {Sadeghi}}]{BitaghsirFadafan:2010zh}%
  \BibitemOpen
  \bibfield  {author} {\bibinfo {author} {\bibfnamefont {K.}~\bibnamefont
  {Bitaghsir~Fadafan}}, \bibinfo {author} {\bibfnamefont {B.}~\bibnamefont
  {Pourhassan}}, \ and\ \bibinfo {author} {\bibfnamefont {J.}~\bibnamefont
  {Sadeghi}},\ }\href {\doibase 10.1140/epjc/s10052-011-1785-0} {\bibfield
  {journal} {\bibinfo  {journal} {Eur. Phys. J. C}\ }\textbf {\bibinfo {volume}
  {71}},\ \bibinfo {pages} {1785} (\bibinfo {year} {2011})},\ \Eprint
  {http://arxiv.org/abs/1005.1368} {arXiv:1005.1368 [hep-th]} \BibitemShut
  {NoStop}%
\bibitem [{\citenamefont {Li}\ \emph {et~al.}(2014)\citenamefont {Li},
  \citenamefont {Liao},\ and\ \citenamefont {Huang}}]{Li:2014hja}%
  \BibitemOpen
  \bibfield  {author} {\bibinfo {author} {\bibfnamefont {D.}~\bibnamefont
  {Li}}, \bibinfo {author} {\bibfnamefont {J.}~\bibnamefont {Liao}}, \ and\
  \bibinfo {author} {\bibfnamefont {M.}~\bibnamefont {Huang}},\ }\href
  {\doibase 10.1103/PhysRevD.89.126006} {\bibfield  {journal} {\bibinfo
  {journal} {Phys. Rev. D}\ }\textbf {\bibinfo {volume} {89}},\ \bibinfo
  {pages} {126006} (\bibinfo {year} {2014})},\ \Eprint
  {http://arxiv.org/abs/1401.2035} {arXiv:1401.2035 [hep-ph]} \BibitemShut
  {NoStop}%
\bibitem [{\citenamefont {Zhang}\ and\ \citenamefont
  {Zhu}(2019)}]{Zhang:2019cxu}%
  \BibitemOpen
  \bibfield  {author} {\bibinfo {author} {\bibfnamefont {Z.-q.}\ \bibnamefont
  {Zhang}}\ and\ \bibinfo {author} {\bibfnamefont {X.}~\bibnamefont {Zhu}},\
  }\href {\doibase 10.1140/epjc/s10052-019-6579-9} {\bibfield  {journal}
  {\bibinfo  {journal} {Eur. Phys. J. C}\ }\textbf {\bibinfo {volume} {79}},\
  \bibinfo {pages} {107} (\bibinfo {year} {2019})}\BibitemShut {NoStop}%
\bibitem [{\citenamefont {Liu}\ \emph {et~al.}(2007)\citenamefont {Liu},
  \citenamefont {Rajagopal},\ and\ \citenamefont {Wiedemann}}]{Liu:2006he}%
  \BibitemOpen
  \bibfield  {author} {\bibinfo {author} {\bibfnamefont {H.}~\bibnamefont
  {Liu}}, \bibinfo {author} {\bibfnamefont {K.}~\bibnamefont {Rajagopal}}, \
  and\ \bibinfo {author} {\bibfnamefont {U.~A.}\ \bibnamefont {Wiedemann}},\
  }\href {\doibase 10.1088/1126-6708/2007/03/066} {\bibfield  {journal}
  {\bibinfo  {journal} {JHEP}\ }\textbf {\bibinfo {volume} {03}},\ \bibinfo
  {pages} {066} (\bibinfo {year} {2007})},\ \Eprint
  {http://arxiv.org/abs/hep-ph/0612168} {arXiv:hep-ph/0612168} \BibitemShut
  {NoStop}%
\bibitem [{\citenamefont {Buchel}(2006)}]{Buchel:2006bv}%
  \BibitemOpen
  \bibfield  {author} {\bibinfo {author} {\bibfnamefont {A.}~\bibnamefont
  {Buchel}},\ }\href {\doibase 10.1103/PhysRevD.74.046006} {\bibfield
  {journal} {\bibinfo  {journal} {Phys. Rev. D}\ }\textbf {\bibinfo {volume}
  {74}},\ \bibinfo {pages} {046006} (\bibinfo {year} {2006})},\ \Eprint
  {http://arxiv.org/abs/hep-th/0605178} {arXiv:hep-th/0605178} \BibitemShut
  {NoStop}%
\bibitem [{\citenamefont {Nakano}\ \emph {et~al.}(2007)\citenamefont {Nakano},
  \citenamefont {Teraguchi},\ and\ \citenamefont {Wen}}]{Nakano:2006js}%
  \BibitemOpen
  \bibfield  {author} {\bibinfo {author} {\bibfnamefont {E.}~\bibnamefont
  {Nakano}}, \bibinfo {author} {\bibfnamefont {S.}~\bibnamefont {Teraguchi}}, \
  and\ \bibinfo {author} {\bibfnamefont {W.-Y.}\ \bibnamefont {Wen}},\ }\href
  {\doibase 10.1103/PhysRevD.75.085016} {\bibfield  {journal} {\bibinfo
  {journal} {Phys. Rev. D}\ }\textbf {\bibinfo {volume} {75}},\ \bibinfo
  {pages} {085016} (\bibinfo {year} {2007})},\ \Eprint
  {http://arxiv.org/abs/hep-ph/0608274} {arXiv:hep-ph/0608274} \BibitemShut
  {NoStop}%
\bibitem [{\citenamefont {Zhu}\ \emph {et~al.}(2021)\citenamefont {Zhu},
  \citenamefont {Chen}, \citenamefont {Liu},\ and\ \citenamefont
  {Hou}}]{Zhu:2021nbl}%
  \BibitemOpen
  \bibfield  {author} {\bibinfo {author} {\bibfnamefont {Z.-R.}\ \bibnamefont
  {Zhu}}, \bibinfo {author} {\bibfnamefont {J.-X.}\ \bibnamefont {Chen}},
  \bibinfo {author} {\bibfnamefont {X.-M.}\ \bibnamefont {Liu}}, \ and\
  \bibinfo {author} {\bibfnamefont {D.}~\bibnamefont {Hou}},\ }\href@noop {} {\
   (\bibinfo {year} {2021})},\ \Eprint {http://arxiv.org/abs/2109.02366}
  {arXiv:2109.02366 [hep-ph]} \BibitemShut {NoStop}%
\bibitem [{\citenamefont {Park}\ and\ \citenamefont {Sin}(1998)}]{Park:1998uv}%
  \BibitemOpen
  \bibfield  {author} {\bibinfo {author} {\bibfnamefont {C.}~\bibnamefont
  {Park}}\ and\ \bibinfo {author} {\bibfnamefont {S.-J.}\ \bibnamefont {Sin}},\
  }\href {\doibase 10.1016/S0370-2693(98)01369-0} {\bibfield  {journal}
  {\bibinfo  {journal} {Phys. Lett. B}\ }\textbf {\bibinfo {volume} {444}},\
  \bibinfo {pages} {156} (\bibinfo {year} {1998})},\ \Eprint
  {http://arxiv.org/abs/hep-th/9807156} {arXiv:hep-th/9807156} \BibitemShut
  {NoStop}%
\bibitem [{\citenamefont {Li}\ and\ \citenamefont {Lin}(2018)}]{Li:2017ywp}%
  \BibitemOpen
  \bibfield  {author} {\bibinfo {author} {\bibfnamefont {S.-w.}\ \bibnamefont
  {Li}}\ and\ \bibinfo {author} {\bibfnamefont {S.}~\bibnamefont {Lin}},\
  }\href {\doibase 10.1103/PhysRevD.98.066002} {\bibfield  {journal} {\bibinfo
  {journal} {Phys. Rev. D}\ }\textbf {\bibinfo {volume} {98}},\ \bibinfo
  {pages} {066002} (\bibinfo {year} {2018})},\ \Eprint
  {http://arxiv.org/abs/1711.06365} {arXiv:1711.06365 [hep-th]} \BibitemShut
  {NoStop}%
\bibitem [{\citenamefont {Liu}\ and\ \citenamefont
  {Tseytlin}(1999)}]{Liu:1999fc}%
  \BibitemOpen
  \bibfield  {author} {\bibinfo {author} {\bibfnamefont {H.}~\bibnamefont
  {Liu}}\ and\ \bibinfo {author} {\bibfnamefont {A.~A.}\ \bibnamefont
  {Tseytlin}},\ }\href {\doibase 10.1016/S0550-3213(99)00259-X} {\bibfield
  {journal} {\bibinfo  {journal} {Nucl. Phys. B}\ }\textbf {\bibinfo {volume}
  {553}},\ \bibinfo {pages} {231} (\bibinfo {year} {1999})},\ \Eprint
  {http://arxiv.org/abs/hep-th/9903091} {arXiv:hep-th/9903091} \BibitemShut
  {NoStop}%
\bibitem [{\citenamefont {Zhang}\ \emph {et~al.}(2018)\citenamefont {Zhang},
  \citenamefont {Luo},\ and\ \citenamefont {Hou}}]{Zhang:2018rff}%
  \BibitemOpen
  \bibfield  {author} {\bibinfo {author} {\bibfnamefont {Z.-q.}\ \bibnamefont
  {Zhang}}, \bibinfo {author} {\bibfnamefont {Z.-j.}\ \bibnamefont {Luo}}, \
  and\ \bibinfo {author} {\bibfnamefont {D.-f.}\ \bibnamefont {Hou}},\ }\href
  {\doibase 10.1016/j.nuclphysa.2018.03.004} {\bibfield  {journal} {\bibinfo
  {journal} {Nucl. Phys. A}\ }\textbf {\bibinfo {volume} {974}},\ \bibinfo
  {pages} {1} (\bibinfo {year} {2018})},\ \Eprint
  {http://arxiv.org/abs/1804.05517} {arXiv:1804.05517 [hep-th]} \BibitemShut
  {NoStop}%
\bibitem [{\citenamefont {Zhang}\ \emph
  {et~al.}(2017{\natexlab{b}})\citenamefont {Zhang}, \citenamefont {Hou},\ and\
  \citenamefont {Chen}}]{Zhang:2017aoc}%
  \BibitemOpen
  \bibfield  {author} {\bibinfo {author} {\bibfnamefont {Z.-q.}\ \bibnamefont
  {Zhang}}, \bibinfo {author} {\bibfnamefont {D.-f.}\ \bibnamefont {Hou}}, \
  and\ \bibinfo {author} {\bibfnamefont {G.}~\bibnamefont {Chen}},\ }\href
  {\doibase 10.1088/1361-6471/aa8daa} {\bibfield  {journal} {\bibinfo
  {journal} {J. Phys. G}\ }\textbf {\bibinfo {volume} {44}},\ \bibinfo {pages}
  {115001} (\bibinfo {year} {2017}{\natexlab{b}})},\ \Eprint
  {http://arxiv.org/abs/1710.06579} {arXiv:1710.06579 [hep-th]} \BibitemShut
  {NoStop}%
\bibitem [{\citenamefont {Shahkarami}\ \emph {et~al.}(2018)\citenamefont
  {Shahkarami}, \citenamefont {Dehghani},\ and\ \citenamefont
  {Dehghani}}]{Shahkarami:2015qff}%
  \BibitemOpen
  \bibfield  {author} {\bibinfo {author} {\bibfnamefont {L.}~\bibnamefont
  {Shahkarami}}, \bibinfo {author} {\bibfnamefont {M.}~\bibnamefont
  {Dehghani}}, \ and\ \bibinfo {author} {\bibfnamefont {P.}~\bibnamefont
  {Dehghani}},\ }\href {\doibase 10.1103/PhysRevD.97.046013} {\bibfield
  {journal} {\bibinfo  {journal} {Phys. Rev. D}\ }\textbf {\bibinfo {volume}
  {97}},\ \bibinfo {pages} {046013} (\bibinfo {year} {2018})},\ \Eprint
  {http://arxiv.org/abs/1511.07986} {arXiv:1511.07986 [hep-th]} \BibitemShut
  {NoStop}%
\bibitem [{\citenamefont {Zhang}\ \emph {et~al.}(2016)\citenamefont {Zhang},
  \citenamefont {Hou},\ and\ \citenamefont {Chen}}]{Zhang:2016fdk}%
  \BibitemOpen
  \bibfield  {author} {\bibinfo {author} {\bibfnamefont {Z.-q.}\ \bibnamefont
  {Zhang}}, \bibinfo {author} {\bibfnamefont {D.-f.}\ \bibnamefont {Hou}}, \
  and\ \bibinfo {author} {\bibfnamefont {G.}~\bibnamefont {Chen}},\ }\href
  {\doibase 10.1140/epja/i2016-16357-9} {\bibfield  {journal} {\bibinfo
  {journal} {Eur. Phys. J. A}\ }\textbf {\bibinfo {volume} {52}},\ \bibinfo
  {pages} {357} (\bibinfo {year} {2016})},\ \Eprint
  {http://arxiv.org/abs/1607.03985} {arXiv:1607.03985 [hep-ph]} \BibitemShut
  {NoStop}%
\bibitem [{\citenamefont {Adamczyk}\ \emph {et~al.}(2017)\citenamefont
  {Adamczyk} \emph {et~al.}}]{STAR:2017ckg}%
  \BibitemOpen
  \bibfield  {author} {\bibinfo {author} {\bibfnamefont {L.}~\bibnamefont
  {Adamczyk}} \emph {et~al.} (\bibinfo {collaboration} {STAR}),\ }\href
  {\doibase 10.1038/nature23004} {\bibfield  {journal} {\bibinfo  {journal}
  {Nature}\ }\textbf {\bibinfo {volume} {548}},\ \bibinfo {pages} {62}
  (\bibinfo {year} {2017})},\ \Eprint {http://arxiv.org/abs/1701.06657}
  {arXiv:1701.06657 [nucl-ex]} \BibitemShut {NoStop}%
\bibitem [{\citenamefont {Abelev}\ \emph {et~al.}(2007)\citenamefont {Abelev}
  \emph {et~al.}}]{STAR:2007ccu}%
  \BibitemOpen
  \bibfield  {author} {\bibinfo {author} {\bibfnamefont {B.~I.}\ \bibnamefont
  {Abelev}} \emph {et~al.} (\bibinfo {collaboration} {STAR}),\ }\href {\doibase
  10.1103/PhysRevC.76.024915} {\bibfield  {journal} {\bibinfo  {journal} {Phys.
  Rev. C}\ }\textbf {\bibinfo {volume} {76}},\ \bibinfo {pages} {024915}
  (\bibinfo {year} {2007})},\ \bibinfo {note} {[Erratum: Phys.Rev.C 95, 039906
  (2017)]},\ \Eprint {http://arxiv.org/abs/0705.1691} {arXiv:0705.1691
  [nucl-ex]} \BibitemShut {NoStop}%
\bibitem [{\citenamefont {Fujimoto}\ \emph {et~al.}(2021)\citenamefont
  {Fujimoto}, \citenamefont {Fukushima},\ and\ \citenamefont
  {Hidaka}}]{Fujimoto:2021xix}%
  \BibitemOpen
  \bibfield  {author} {\bibinfo {author} {\bibfnamefont {Y.}~\bibnamefont
  {Fujimoto}}, \bibinfo {author} {\bibfnamefont {K.}~\bibnamefont {Fukushima}},
  \ and\ \bibinfo {author} {\bibfnamefont {Y.}~\bibnamefont {Hidaka}},\ }\href
  {\doibase 10.1016/j.physletb.2021.136184} {\bibfield  {journal} {\bibinfo
  {journal} {Phys. Lett. B}\ }\textbf {\bibinfo {volume} {816}},\ \bibinfo
  {pages} {136184} (\bibinfo {year} {2021})},\ \Eprint
  {http://arxiv.org/abs/2101.09173} {arXiv:2101.09173 [hep-ph]} \BibitemShut
  {NoStop}%
\bibitem [{\citenamefont {Chernodub}(2021)}]{Chernodub:2020qah}%
  \BibitemOpen
  \bibfield  {author} {\bibinfo {author} {\bibfnamefont {M.~N.}\ \bibnamefont
  {Chernodub}},\ }\href {\doibase 10.1103/PhysRevD.103.054027} {\bibfield
  {journal} {\bibinfo  {journal} {Phys. Rev. D}\ }\textbf {\bibinfo {volume}
  {103}},\ \bibinfo {pages} {054027} (\bibinfo {year} {2021})},\ \Eprint
  {http://arxiv.org/abs/2012.04924} {arXiv:2012.04924 [hep-ph]} \BibitemShut
  {NoStop}%
\bibitem [{\citenamefont {Chernodub}\ and\ \citenamefont
  {Gongyo}(2017)}]{Chernodub:2016kxh}%
  \BibitemOpen
  \bibfield  {author} {\bibinfo {author} {\bibfnamefont {M.~N.}\ \bibnamefont
  {Chernodub}}\ and\ \bibinfo {author} {\bibfnamefont {S.}~\bibnamefont
  {Gongyo}},\ }\href {\doibase 10.1007/JHEP01(2017)136} {\bibfield  {journal}
  {\bibinfo  {journal} {JHEP}\ }\textbf {\bibinfo {volume} {01}},\ \bibinfo
  {pages} {136} (\bibinfo {year} {2017})},\ \Eprint
  {http://arxiv.org/abs/1611.02598} {arXiv:1611.02598 [hep-th]} \BibitemShut
  {NoStop}%
\bibitem [{\citenamefont {Zhou}\ \emph {et~al.}(2021)\citenamefont {Zhou},
  \citenamefont {Chen}, \citenamefont {Zhao},\ and\ \citenamefont
  {Ping}}]{Zhou:2021sdy}%
  \BibitemOpen
  \bibfield  {author} {\bibinfo {author} {\bibfnamefont {J.}~\bibnamefont
  {Zhou}}, \bibinfo {author} {\bibfnamefont {X.}~\bibnamefont {Chen}}, \bibinfo
  {author} {\bibfnamefont {Y.-Q.}\ \bibnamefont {Zhao}}, \ and\ \bibinfo
  {author} {\bibfnamefont {J.}~\bibnamefont {Ping}},\ }\href {\doibase
  10.1103/PhysRevD.102.126029} {\bibfield  {journal} {\bibinfo  {journal}
  {Phys. Rev. D}\ }\textbf {\bibinfo {volume} {102}},\ \bibinfo {pages}
  {126029} (\bibinfo {year} {2021})}\BibitemShut {NoStop}%
\bibitem [{\citenamefont {Golubtsova}\ \emph {et~al.}(2021)\citenamefont
  {Golubtsova}, \citenamefont {Gourgoulhon},\ and\ \citenamefont
  {Usova}}]{Golubtsova:2021agl}%
  \BibitemOpen
  \bibfield  {author} {\bibinfo {author} {\bibfnamefont {A.~A.}\ \bibnamefont
  {Golubtsova}}, \bibinfo {author} {\bibfnamefont {E.}~\bibnamefont
  {Gourgoulhon}}, \ and\ \bibinfo {author} {\bibfnamefont {M.~K.}\ \bibnamefont
  {Usova}},\ }\href@noop {} {\  (\bibinfo {year} {2021})},\ \Eprint
  {http://arxiv.org/abs/2107.11672} {arXiv:2107.11672 [hep-th]} \BibitemShut
  {NoStop}%
\bibitem [{\citenamefont {McInnes}(2018)}]{McInnes:2018ibt}%
  \BibitemOpen
  \bibfield  {author} {\bibinfo {author} {\bibfnamefont {B.}~\bibnamefont
  {McInnes}},\ }\href@noop {} {\  (\bibinfo {year} {2018})},\ \Eprint
  {http://arxiv.org/abs/1808.00648} {arXiv:1808.00648 [hep-th]} \BibitemShut
  {NoStop}%
\bibitem [{\citenamefont {McInnes}(2020)}]{McInnes:2018mwj}%
  \BibitemOpen
  \bibfield  {author} {\bibinfo {author} {\bibfnamefont {B.}~\bibnamefont
  {McInnes}},\ }\href {\doibase 10.1016/j.nuclphysb.2020.114951} {\bibfield
  {journal} {\bibinfo  {journal} {Nucl. Phys. B}\ }\textbf {\bibinfo {volume}
  {953}},\ \bibinfo {pages} {114951} (\bibinfo {year} {2020})},\ \Eprint
  {http://arxiv.org/abs/1812.07146} {arXiv:1812.07146 [hep-th]} \BibitemShut
  {NoStop}%
\bibitem [{\citenamefont {Adam}\ \emph {et~al.}(2018)\citenamefont {Adam} \emph
  {et~al.}}]{STAR:2018gyt}%
  \BibitemOpen
  \bibfield  {author} {\bibinfo {author} {\bibfnamefont {J.}~\bibnamefont
  {Adam}} \emph {et~al.} (\bibinfo {collaboration} {STAR}),\ }\href {\doibase
  10.1103/PhysRevC.98.014910} {\bibfield  {journal} {\bibinfo  {journal} {Phys.
  Rev. C}\ }\textbf {\bibinfo {volume} {98}},\ \bibinfo {pages} {014910}
  (\bibinfo {year} {2018})},\ \Eprint {http://arxiv.org/abs/1805.04400}
  {arXiv:1805.04400 [nucl-ex]} \BibitemShut {NoStop}%
\bibitem [{\citenamefont {Ebihara}\ \emph {et~al.}(2017)\citenamefont
  {Ebihara}, \citenamefont {Fukushima},\ and\ \citenamefont
  {Mameda}}]{Ebihara:2016fwa}%
  \BibitemOpen
  \bibfield  {author} {\bibinfo {author} {\bibfnamefont {S.}~\bibnamefont
  {Ebihara}}, \bibinfo {author} {\bibfnamefont {K.}~\bibnamefont {Fukushima}},
  \ and\ \bibinfo {author} {\bibfnamefont {K.}~\bibnamefont {Mameda}},\ }\href
  {\doibase 10.1016/j.physletb.2016.11.010} {\bibfield  {journal} {\bibinfo
  {journal} {Phys. Lett. B}\ }\textbf {\bibinfo {volume} {764}},\ \bibinfo
  {pages} {94} (\bibinfo {year} {2017})},\ \Eprint
  {http://arxiv.org/abs/1608.00336} {arXiv:1608.00336 [hep-ph]} \BibitemShut
  {NoStop}%
\bibitem [{\citenamefont {Becattini}\ and\ \citenamefont
  {Lisa}(2020)}]{Becattini:2020ngo}%
  \BibitemOpen
  \bibfield  {author} {\bibinfo {author} {\bibfnamefont {F.}~\bibnamefont
  {Becattini}}\ and\ \bibinfo {author} {\bibfnamefont {M.~A.}\ \bibnamefont
  {Lisa}},\ }\href {\doibase 10.1146/annurev-nucl-021920-095245} {\bibfield
  {journal} {\bibinfo  {journal} {Ann. Rev. Nucl. Part. Sci.}\ }\textbf
  {\bibinfo {volume} {70}},\ \bibinfo {pages} {395} (\bibinfo {year} {2020})},\
  \Eprint {http://arxiv.org/abs/2003.03640} {arXiv:2003.03640 [nucl-ex]}
  \BibitemShut {NoStop}%
\bibitem [{\citenamefont {McInnes}(2016)}]{McInnes:2016dwk}%
  \BibitemOpen
  \bibfield  {author} {\bibinfo {author} {\bibfnamefont {B.}~\bibnamefont
  {McInnes}},\ }\href {\doibase 10.1016/j.nuclphysb.2016.08.001} {\bibfield
  {journal} {\bibinfo  {journal} {Nucl. Phys. B}\ }\textbf {\bibinfo {volume}
  {911}},\ \bibinfo {pages} {173} (\bibinfo {year} {2016})},\ \Eprint
  {http://arxiv.org/abs/1604.03669} {arXiv:1604.03669 [hep-th]} \BibitemShut
  {NoStop}%
\bibitem [{\citenamefont {Bravo~Gaete}\ \emph {et~al.}(2017)\citenamefont
  {Bravo~Gaete}, \citenamefont {Guajardo},\ and\ \citenamefont
  {Hassaine}}]{BravoGaete:2017dso}%
  \BibitemOpen
  \bibfield  {author} {\bibinfo {author} {\bibfnamefont {M.}~\bibnamefont
  {Bravo~Gaete}}, \bibinfo {author} {\bibfnamefont {L.}~\bibnamefont
  {Guajardo}}, \ and\ \bibinfo {author} {\bibfnamefont {M.}~\bibnamefont
  {Hassaine}},\ }\href {\doibase 10.1007/JHEP04(2017)092} {\bibfield  {journal}
  {\bibinfo  {journal} {JHEP}\ }\textbf {\bibinfo {volume} {04}},\ \bibinfo
  {pages} {092} (\bibinfo {year} {2017})},\ \Eprint
  {http://arxiv.org/abs/1702.02416} {arXiv:1702.02416 [hep-th]} \BibitemShut
  {NoStop}%
\bibitem [{\citenamefont {Erices}\ and\ \citenamefont
  {Martinez}(2018)}]{Erices:2017izj}%
  \BibitemOpen
  \bibfield  {author} {\bibinfo {author} {\bibfnamefont {C.}~\bibnamefont
  {Erices}}\ and\ \bibinfo {author} {\bibfnamefont {C.}~\bibnamefont
  {Martinez}},\ }\href {\doibase 10.1103/PhysRevD.97.024034} {\bibfield
  {journal} {\bibinfo  {journal} {Phys. Rev. D}\ }\textbf {\bibinfo {volume}
  {97}},\ \bibinfo {pages} {024034} (\bibinfo {year} {2018})},\ \Eprint
  {http://arxiv.org/abs/1707.03483} {arXiv:1707.03483 [hep-th]} \BibitemShut
  {NoStop}%
\bibitem [{\citenamefont {Chen}\ \emph {et~al.}(2020)\citenamefont {Chen},
  \citenamefont {Zhang}, \citenamefont {Li}, \citenamefont {Hou},\ and\
  \citenamefont {Huang}}]{Chen:2020ath}%
  \BibitemOpen
  \bibfield  {author} {\bibinfo {author} {\bibfnamefont {X.}~\bibnamefont
  {Chen}}, \bibinfo {author} {\bibfnamefont {L.}~\bibnamefont {Zhang}},
  \bibinfo {author} {\bibfnamefont {D.}~\bibnamefont {Li}}, \bibinfo {author}
  {\bibfnamefont {D.}~\bibnamefont {Hou}}, \ and\ \bibinfo {author}
  {\bibfnamefont {M.}~\bibnamefont {Huang}},\ }\href@noop {} {\  (\bibinfo
  {year} {2020})},\ \Eprint {http://arxiv.org/abs/2010.14478} {arXiv:2010.14478
  [hep-ph]} \BibitemShut {NoStop}%
\bibitem [{\citenamefont {Gibbons}\ \emph {et~al.}(1996)\citenamefont
  {Gibbons}, \citenamefont {Green},\ and\ \citenamefont
  {Perry}}]{Gibbons:1995vg}%
  \BibitemOpen
  \bibfield  {author} {\bibinfo {author} {\bibfnamefont {G.~W.}\ \bibnamefont
  {Gibbons}}, \bibinfo {author} {\bibfnamefont {M.~B.}\ \bibnamefont {Green}},
  \ and\ \bibinfo {author} {\bibfnamefont {M.~J.}\ \bibnamefont {Perry}},\
  }\href {\doibase 10.1016/0370-2693(95)01565-5} {\bibfield  {journal}
  {\bibinfo  {journal} {Phys. Lett. B}\ }\textbf {\bibinfo {volume} {370}},\
  \bibinfo {pages} {37} (\bibinfo {year} {1996})},\ \Eprint
  {http://arxiv.org/abs/hep-th/9511080} {arXiv:hep-th/9511080} \BibitemShut
  {NoStop}%
\bibitem [{\citenamefont {Kehagias}\ and\ \citenamefont
  {Sfetsos}(1999)}]{Kehagias:1999iy}%
  \BibitemOpen
  \bibfield  {author} {\bibinfo {author} {\bibfnamefont {A.}~\bibnamefont
  {Kehagias}}\ and\ \bibinfo {author} {\bibfnamefont {K.}~\bibnamefont
  {Sfetsos}},\ }\href {\doibase 10.1016/S0370-2693(99)00431-1} {\bibfield
  {journal} {\bibinfo  {journal} {Phys. Lett. B}\ }\textbf {\bibinfo {volume}
  {456}},\ \bibinfo {pages} {22} (\bibinfo {year} {1999})},\ \Eprint
  {http://arxiv.org/abs/hep-th/9903109} {arXiv:hep-th/9903109} \BibitemShut
  {NoStop}%
\bibitem [{\citenamefont {Li}\ \emph {et~al.}(2021)\citenamefont {Li},
  \citenamefont {Luo},\ and\ \citenamefont {Tan}}]{Li:2021vve}%
  \BibitemOpen
  \bibfield  {author} {\bibinfo {author} {\bibfnamefont {S.-w.}\ \bibnamefont
  {Li}}, \bibinfo {author} {\bibfnamefont {S.-k.}\ \bibnamefont {Luo}}, \ and\
  \bibinfo {author} {\bibfnamefont {M.-z.}\ \bibnamefont {Tan}},\ }\href
  {\doibase 10.1103/PhysRevD.104.066008} {\bibfield  {journal} {\bibinfo
  {journal} {Phys. Rev. D}\ }\textbf {\bibinfo {volume} {104}},\ \bibinfo
  {pages} {066008} (\bibinfo {year} {2021})},\ \Eprint
  {http://arxiv.org/abs/2106.04038} {arXiv:2106.04038 [hep-th]} \BibitemShut
  {NoStop}%
\bibitem [{\citenamefont {Gwak}\ \emph {et~al.}(2012)\citenamefont {Gwak},
  \citenamefont {Kim}, \citenamefont {Lee}, \citenamefont {Seo},\ and\
  \citenamefont {Sin}}]{Gwak:2012ht}%
  \BibitemOpen
  \bibfield  {author} {\bibinfo {author} {\bibfnamefont {B.}~\bibnamefont
  {Gwak}}, \bibinfo {author} {\bibfnamefont {M.}~\bibnamefont {Kim}}, \bibinfo
  {author} {\bibfnamefont {B.-H.}\ \bibnamefont {Lee}}, \bibinfo {author}
  {\bibfnamefont {Y.}~\bibnamefont {Seo}}, \ and\ \bibinfo {author}
  {\bibfnamefont {S.-J.}\ \bibnamefont {Sin}},\ }\href {\doibase
  10.1103/PhysRevD.86.026010} {\bibfield  {journal} {\bibinfo  {journal} {Phys.
  Rev. D}\ }\textbf {\bibinfo {volume} {86}},\ \bibinfo {pages} {026010}
  (\bibinfo {year} {2012})},\ \Eprint {http://arxiv.org/abs/1203.4883}
  {arXiv:1203.4883 [hep-th]} \BibitemShut {NoStop}%
\bibitem [{\citenamefont {Shahkarami}\ and\ \citenamefont
  {Charmchi}(2019)}]{Shahkarami:2019zax}%
  \BibitemOpen
  \bibfield  {author} {\bibinfo {author} {\bibfnamefont {L.}~\bibnamefont
  {Shahkarami}}\ and\ \bibinfo {author} {\bibfnamefont {F.}~\bibnamefont
  {Charmchi}},\ }\href {\doibase 10.1140/epjc/s10052-019-6765-9} {\bibfield
  {journal} {\bibinfo  {journal} {Eur. Phys. J. C}\ }\textbf {\bibinfo {volume}
  {79}},\ \bibinfo {pages} {343} (\bibinfo {year} {2019})},\ \Eprint
  {http://arxiv.org/abs/1904.09806} {arXiv:1904.09806 [hep-th]} \BibitemShut
  {NoStop}%
\bibitem [{\citenamefont {Gubser}(2007)}]{Gubser:2006qh}%
  \BibitemOpen
  \bibfield  {author} {\bibinfo {author} {\bibfnamefont {S.~S.}\ \bibnamefont
  {Gubser}},\ }\href {\doibase 10.1103/PhysRevD.76.126003} {\bibfield
  {journal} {\bibinfo  {journal} {Phys. Rev. D}\ }\textbf {\bibinfo {volume}
  {76}},\ \bibinfo {pages} {126003} (\bibinfo {year} {2007})},\ \Eprint
  {http://arxiv.org/abs/hep-th/0611272} {arXiv:hep-th/0611272} \BibitemShut
  {NoStop}%
\bibitem [{\citenamefont {Edelstein}\ and\ \citenamefont
  {Salgado}(2008)}]{Edelstein:2008cp}%
  \BibitemOpen
  \bibfield  {author} {\bibinfo {author} {\bibfnamefont {J.~D.}\ \bibnamefont
  {Edelstein}}\ and\ \bibinfo {author} {\bibfnamefont {C.~A.}\ \bibnamefont
  {Salgado}},\ }\href {\doibase 10.1063/1.2972007} {\bibfield  {journal}
  {\bibinfo  {journal} {AIP Conf. Proc.}\ }\textbf {\bibinfo {volume} {1031}},\
  \bibinfo {pages} {207} (\bibinfo {year} {2008})},\ \Eprint
  {http://arxiv.org/abs/0805.4515} {arXiv:0805.4515 [hep-th]} \BibitemShut
  {NoStop}%
\end{thebibliography}%
\end{document}